\documentclass[cits]{PoS_mod}
\pdfoutput=1
\usepackage{times}
\usepackage[T1]{fontenc}
\usepackage{graphicx}
\usepackage{amsmath}

\newcommand{\MeV}{~\mathrm{MeV}}

\newcommand{\fm}{~\mathrm{fm}}

\newcommand{\SU}{\mathrm{SU}}

\newcommand{\appropto}{\mathrel{\vcenter{
  \offinterlineskip\halign{\hfil$##$\cr
    \propto\cr\noalign{\kern2pt}\sim\cr\noalign{\kern-2pt}}}}}

\title{Towards the continuum limit with improved Wilson fermions employing open boundary conditions}

\ShortTitle{Towards the continuum limit with improved Wilson fermions}

\speakershort{Gunnar S.~Bali and Wolfgang S\"oldner}
\author{Gunnar S.~Bali$^*$,$^{ab}$ Sara Collins,$^a$
Fabian Hutzler,$^a$ Meinulf G\"ockeler,$^a$\newline
Andreas Sch\"afer,$^a$
Enno E.~Scholz,$^a$ Jakob Simeth,$^a$ \speaker{Wolfgang S\"oldner},$^a$\newline
Andr\'e Sternbeck$^c$
and Thomas Wurm$^a$\newline
(RQCD Collaboration)\thanks{
This work was supported by DFG SFB/TRR 55.
We acknowledge the Gauss Centre for Supercomputing e.V.~for granting
computer time on SuperMUC at Leibniz Supercomputing Centre Munich
and JUQUEEN at J\"ulich Supercomputing Centre as well as
PRACE for providing time on Fermi at
CINECA Bologna and on SuperMUC. Additional ensembles were generated
on the SFB/TRR~55 QPACE computer and some of the analysis was carried out
on the SFB/TRR~55 QPACE~2~\cite{Arts:2015jia}
Xeon-PHI installation in Regensburg. The ensembles were generated
using {\sc openQCD}~\cite{Luscher:2012av} and
(on QPACE) {\sc BQCD}~\cite{Nakamura:2010qh}.
We used the {\sc CHROMA}~\cite{Edwards:2004sx} software package along with the
{\sc LibHadronAnalysis} library and the multigrid solver implementation
of Ref.~\cite{Heybrock:2015kpy} (see also
Ref.~\cite{Frommer:2013fsa})
to generate hadronic two-point functions.
We thank Benjamin Gl\"a\ss{}le, Piotr Korcyl and Daniel Richtmann for
code development, discussions and software support.
Last but not least we thank all our CLS colleagues who made this possible.}

\\
\llap{$^a$}Institut f\"ur Theoretische Physik, Universit\"at Regensburg\\
D-93040 Regensburg, Germany\\
\llap{$^b$}Tata Institute of Fundamental Research, Homi Bhabha Road\\
Mumbai 400005, India\\
\llap{$^c$}Theoretisch-Physikalisches Institut, Friedrich-Schiller-Universit\"at Jena\\
07743 Jena, Germany\\
E-mail: \email{gunnar.bali@ur.de}\\
\hspace*{1.2cm}\email{wolfgang.soeldner@physik.uni-regensburg.de}}

\abstract{We present selected results obtained by RQCD
from simulations of $N_f=2+1$ flavours of non-perturbatively
$\mathcal{O}(a)$ improved Wilson fermions, employing open boundary
conditions in time. The ensembles were created within the CLS
(Coordinated Lattice Simulations) effort
at five different values of the lattice spacing, ranging from 0.085~fm
down to below 0.04~fm. Many quark mass combinations were realized,
in particular along lines where the sum of the bare quark masses was kept
fixed as well as trajectories of an approximately physical renormalized
strange quark mass. Several key observables, including meson and baryon
masses and the axial charge of the nucleon have been computed, and
preliminary results are presented here. In some cases an accurate and
controlled extrapolation to the continuum limit has become possible.}

\FullConference{34th annual International Symposium on Lattice Field Theory\\
		24-30 July 2016\\
		University of Southampton, UK}

\begin{document}
\section{Introduction}
Present-day lattice simulations are typically carried out employing highly
optimized algorithms and utilizing large amounts of computing power.
With increased statistical precision it has become compulsory to
control all systematics including effects due to the finite simulation
volume, unphysical values of the quark masses and the lattice
cut-off $a>0$. The results presented here have all been obtained
on volumes with a linear spatial extent $L\gtrsim\max\{
4m_{\rm PS}^{-1},2\fm\}$, where $m_{\rm PS}$ denotes the mass
of the lightest pseudoscalar meson. Moreover, as will be described
in more detail below, two different trajectories 
were used to safely extrapolate to the physical point
in the quark mass plane.

Taking the continuum
limit is at least equally important. To this end we utilize
$\mathcal{O}(a)$ improvement, however, the finest two of our
five lattice spacings are in the regime where
topological freezing~\cite{DelDebbio:2002xa, Bernard:2003gq, Schaefer:2010hu} becomes a serious problem. This sets in, dependent on the
details of the simulation, between $a^{-1}=3\,\text{GeV}$ and
$4\,\text{GeV}$~\cite{Bruno:2014ova} while our finest lattice corresponds to
$a^{-1}\approx 5\,\text{GeV}$.
Large autocorrelation times for quantities which couple to topological
modes and non-ergodicity of the simulation can be avoided by
employing open boundary conditions (OBC)~\cite{Luscher:2011kk}, that
allow objects carrying topological charge to flow into and out of
the simulation volume. Here, we employ OBC in time and
periodic boundary conditions (PBC) in space.
We remark that the computational overhead, due to the use of OBC
is quite moderate~\cite{Luscher:2012av} and even
more so if one considers the slowing down of simulations with conventional
boundary conditions with at least a large power of $a^{-1}$ if not
exponentially.

The CLS group (Coordinated Lattice Simulations)~\cite{Bruno:2014jqa} has started a
major effort to generate $N_f=2+1$ gauge ensembles employing
non-perturbatively improved Wilson fermions with OBC
aiming at removing all the above mentioned systematics and in particular
enabling controlled continuum limit extrapolations, utilizing at least
five lattice spacings. Some of the available ensembles
are detailed in Refs.~\cite{Bruno:2014jqa,Bali:2016umi}.

Here we present preliminary results on the octet and decuplet baryon spectra
(in Sec.~\ref{sec:baryon}), discuss scale setting (in Sec.~\ref{sec:scale})
and attempt a first continuum limit extrapolation at the $m_u=m_d=m_s$
point of the ratio of the octet over the decuplet baryon mass as well
as of the axial charge (Sec.~\ref{sec:cont}).
Prior to this, we present some details about the simulation strategy
and the fitting procedure~(Sec.~\ref{sec:sim}~--~\ref{sec:meas}).

\begin{figure}[t]
    \centerline{\includegraphics[width = .55\textwidth]{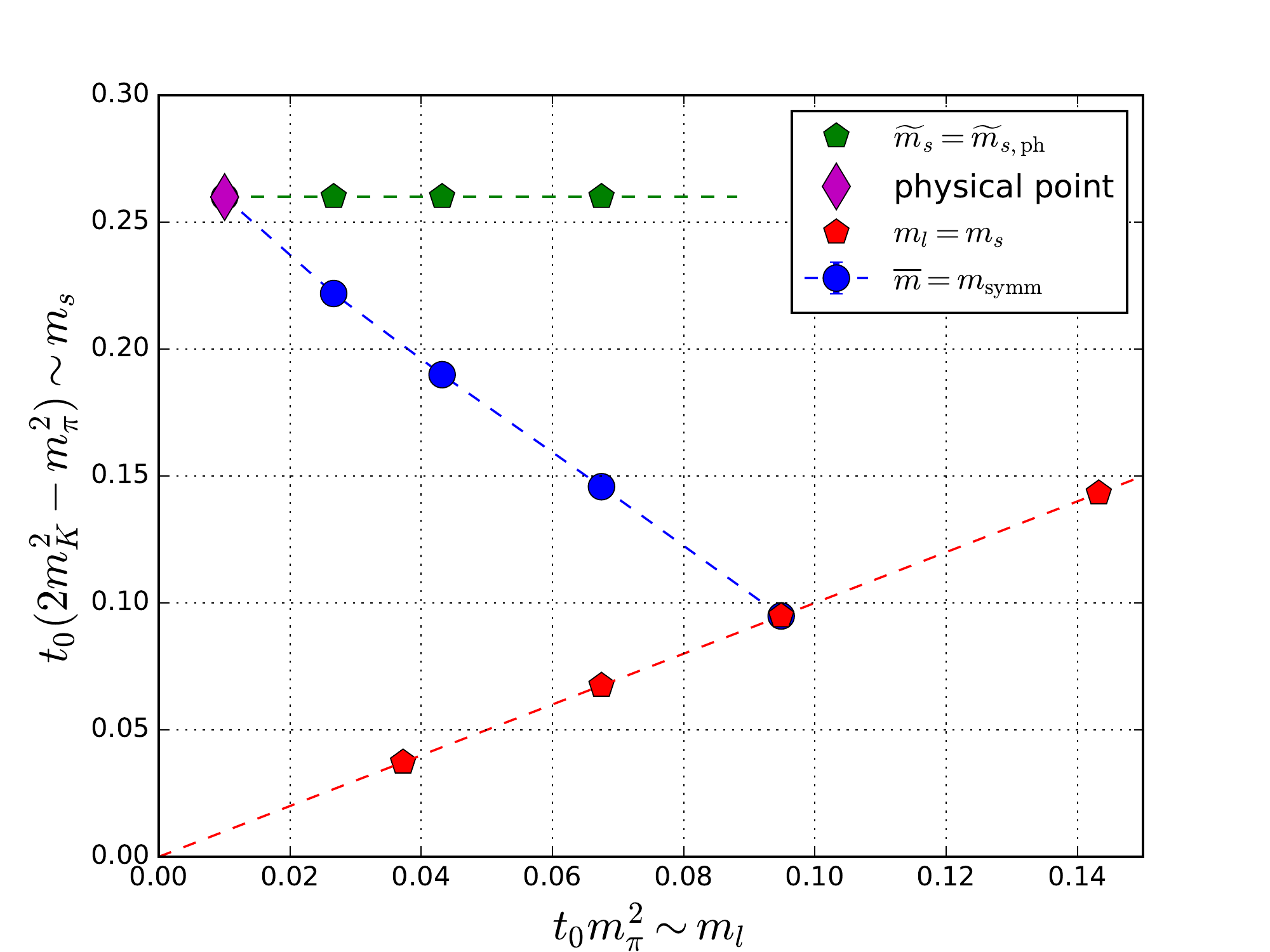}}
    \caption{\label{fig:sim} General simulation strategy in the quark mass plane (strange quark vs.\ light quark mass). The red line represents the flavour
symmetric trajectory,
    the blue and green lines show the trajectories along a constant average quark mass and a constant strange quark mass, respectively.}
\end{figure} 
\section{Overview of the simulation strategy\label{sec:sim}}
We use the non-perturbatively $\mathcal{O}(a)$-improved Wilson fermion action
with tree-level Symanzik improved gauge action and $2+1$ flavours
of degenerate light quarks of mass $m_{\ell}=m_u=m_d$
and a strange quark of mass $m_s$. Our simulation strategy in
terms of the quark masses is outlined in Fig.~\ref{fig:sim}.
In general, we realize three chiral trajectories for each value
of the inverse gauge coupling $\beta=6/g^2$:

(1) Fixed average quark mass, $\overline m = m_\mathrm{symm}$:\\
We keep the sum of bare quark masses constant:
$3\overline{m}=2m_{\ell} + m_{s} = \text{const.}$
This is equivalent to
$2/\kappa_{\ell} + 1/\kappa_{s}=\text{const.}$, which
means that the sum of renormalized quark masses
$2\widehat{m}_{\ell}+\widehat{m}_s=\text{const.}+\mathcal{O}(a)$.

(2) Fixed strange quark mass, $\widetilde{m}_s =\widetilde{m}_{s,\mathrm{ph}}$, see Ref.~\cite{Bali:2016umi}:\\
The strange axial ward identity (AWI) mass $\widetilde{m}_s$ is kept
fixed resulting in a renormalized strange quark mass
$\widehat{m}_s=\widehat{m}_{s,\mathrm{ph}}$, up to tiny $\mathcal{O}(a)$ effects.

(3) Symmetric line, $m_s = m_{\ell}$:\\
For the joint non-perturbative renormalization programme of
the Mainz group and RQCD (as well as to fix the
$\widetilde{m}_s =\widetilde{m}_{s,\mathrm{ph}}$ trajectory),
additional simulations along the flavour-symmetric line are performed.

At present not all of these lines are available for all of our lattice spacings.
Following the strategy introduced by the QCDSF
collaboration~\cite{Bietenholz:2011qq}, at each $\beta$-value we
first determine the $m_s=m_{\ell}$ point at which
the combination $m_K^2+m_\pi^2/2$ assumes its physical value.
Starting from this point, the $\overline{m}=m_{\textrm{symm}}$ chiral
trajectory is controlled by just one
parameter and we benefit from the fact that flavour averaged
quantities vary only moderately.

Since the scale $t_0/a^2$~\cite{Luscher:2010iy}
remains almost constant along this line, see
Fig.~\ref{fig:t0} below, in practice we use the dimensionless combination
$\phi_4 = 8 t_0 (m_K^2 + m_\pi^2/2)$
to fix the $m_{\ell}=m_s=\overline{m}$
starting point and $\sum_i 1/\kappa_i$.
Using the values
$m_\pi  = 134.8(3) \MeV$ and $m_K = 494.2(4) \MeV$ of
Ref.~\cite{Aoki:2016frl}, and $\sqrt{8t_0} = 0.4144(59)(37) \fm$
of Ref.~\cite{Borsanyi:2012zs}, one obtains
$\phi_4^{\mathrm phys} |_{m_{ud}=m_s}=1.117(38)$.
Our original target value $\phi_4 |_{m_{\ell}=m_s}=1.15$ was chosen
somewhat larger
to account for the small slope of the chiral extrapolation found 
in preparatory studies at coarse lattice spacings.

The ensembles along the flavour-symmetric trajectory
($m_{\ell}=m_s$) are used for our non-per\-tur\-ba\-tive renormalization programme,
where we work in a massless scheme and therefore
have to extrapolate to $m_s=m_{\ell}=0$.
In addition, we use this trajectory for fixing the 
$\widetilde{m}_s =\widetilde{m}_{s,\mathrm{ph}}$ simulation parameters,
which is non-trivial in the Wilson formulation, due to additive
mass renormalization and resulting differences between defining singlet
and non-singlet quark mass combinations.

In the following we sketch how the $\kappa$-values along the $\widetilde{m}_s =\widetilde{m}_{s,\mathrm{ph}}$ line are determined:
we first parameterize light and
strange AWI masses as functions of the bare quark masses (including
$\mathrm{O}(a)$-improvement terms), combining
data from both trajectories, $\overline m = m_\mathrm{symm}$ and $m_s = m_{\ell}$.
In a second step we determine the ``physical'' point along
the $\overline m = m_\mathrm{symm}$ line as the point where the ratio
$\widetilde{m}_s/\widetilde{m}_{\ell}$ 
takes its physical value
$\widetilde{m}_s/\widetilde{m}_{\ell}=27.46(44)$~\cite{Aoki:2016frl}.
In this way we obtain $\widetilde{m}_{s,\mathrm{ph}}$ and, finally, with
our parametrization at hand, we can predict the $(\kappa_{\ell},\kappa_s)$
pairs for which $\widetilde{m}_s = \widetilde{m}_{s,\mathrm{ph}}$.
More detail can be found in Sec.~\ref{sec:ms}. 
  
\begin{figure}[t]
\centerline{\includegraphics[width = .5\textwidth]{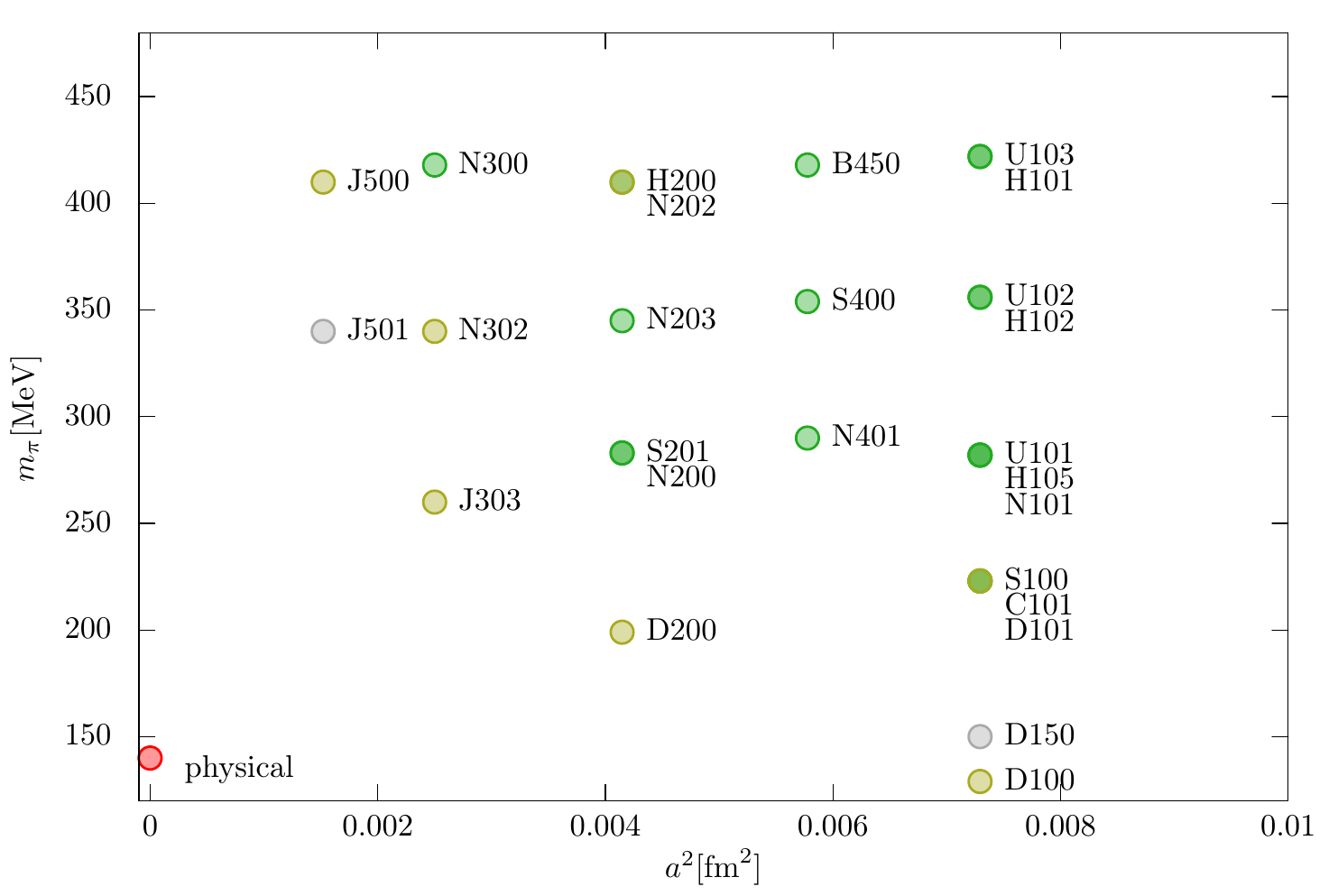} 
\includegraphics[width = .5\textwidth]{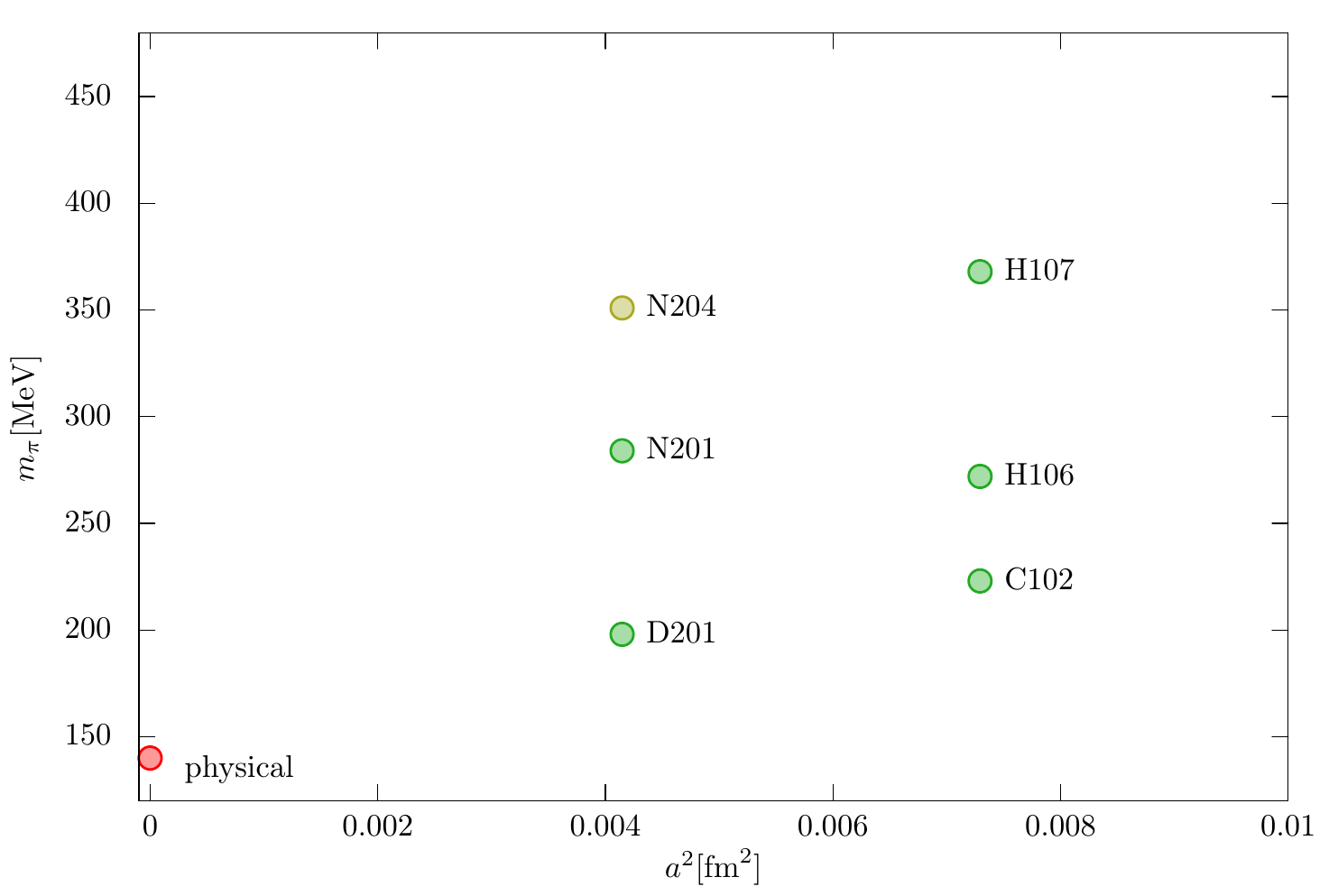}}
\caption{\label{fig:simdetail}CLS ensemble overview, see Refs.~\cite{Bruno:2014jqa,Bali:2016umi}. Ensembles with $\overline m = m_\mathrm{symm}$ (left) and $\widetilde{m}_s = \widetilde{m}_{s,\mathrm{ph}}$ (right). Different letters refer to different volumes/aspect
ratios.}
\end{figure}

\section{Simulation Details\label{sec:detail}}

The gauge field configurations are generated using the {\sc openQCD}
package~\cite{Luscher:2012av}. This contains several algorithmic improvements,
e.g., the Hasenbusch trick, higher order integrators, a multi-level integration
scheme,
a deflated solver~\cite{Luscher:2007es,Frommer:2013fsa} and twisted mass
reweighting: in the light fermion part of the action
a twisted mass
term is introduced in order to push eigenvalues of the Dirac
operator away from zero and, hence, increase the stability of the HMC
simulation. This is then corrected for by reweighting the observables
accordingly. Also with respect to the strange quark action reweighting
is applied to correct for the inaccuracy of the rational approximation
used in the $N_f=1$ part of the HMC simulation.
The reweighting works quite efficiently in practice, more details can
be found in Ref.~\cite{Bruno:2014jqa}.

So far CLS has generated ensembles at five different
values of the inverse gauge coupling:
$\beta = 6/g^2 = 3.4, 3.46, 3.55, 3.7$ and $3.85$.
These couplings correspond to lattice spacings $a$ of
roughly~$0.085\fm, 0.077\fm, 0.064\fm, 0.05\fm$ and $0.04\fm$, respectively,
covering a range of almost a factor five in terms of $a^2$.

Note that the aspect ratios of the generated lattices are usually
larger than two, so that regions that are close to
the classically open boundaries in the time direction can be discarded,
as these are polluted by artefacts related to cut-off effects as well as massive
scalar states propagating into the simulation volume.
All ensembles with OBC have a spatial lattice extent
$L\gtrsim 4m_{\text{PS}}^{-1}$.
Each HMC trajectory has length $\tau=2$ and for the ensembles under
consideration at least 4000 Molecular Dynamic Units (MDUs) have been
generated and often many more.
Along the symmetric line we make use of additional ensembles with
anti-periodic boundary conditions in time. These usually
have less statistics and some of these have $\tau=1$ and
were generated using {\sc BQCD}
software~\cite{Nakamura:2010qh} on the
SFB/TR55 {\sc QPACE} installation. 
For more detail we refer the reader to Refs.~\cite{Bruno:2014jqa,Bali:2016umi}.
An overview over the presently available CLS ensembles is given
in Fig.~\ref{fig:simdetail}.

\section{\label{sec:ms}How to fix the $\widetilde{m}_s =\widetilde{m}_{s,\mathrm{ph}}$ trajectory}
Here we outline the determination of the $\widetilde{m}_s =\widetilde{m}_{s,\mathrm{ph}}$ simulation points.
For simplicity we do not discuss $\mathcal{O}(a)$ improvement, however,
more detail can be found in Ref.~\cite{Bali:2016umi}.
We start with some definitions. The lattice quark masses are given by
$m_j=\left(1/\kappa_j-1/\kappa_{\mathrm{crit}}\right)/(2a)$,
the (averaged) AWI masses are defined as
\begin{equation}
\frac{\widetilde{m}_j+\widetilde{m}_k}{2}
=\widetilde{m}_{jk}=\frac{\partial_4\langle 0|A_4^{jk}|\pi^{jk}\rangle}{
2\langle 0|P^{jk}|\pi^{jk}\rangle}.
\end{equation}
The point along the symmetric line ($m_1=m_2=m_{\ell}=m_s=m_3$) where $\widetilde{m}_{jk}=0$ defines $\kappa_{\mathrm{crit}}$.
A potentially delicate issue is the different renormalization of
flavour-singlet and non-singlet
quark mass combinations. While for the non-singlet combination
\begin{equation}
Z_m(m_s-m_{\ell})=\frac{Z_m}{2a}\left(\frac{1}{\kappa_s}-\frac{1}{\kappa_{\ell}}\right)=\widehat{m}_s-\widehat{m}_{\ell}=\frac{Z_A}{Z_P}2\left(\widetilde{m}_{13}-
\widetilde{m}_{12}\right)
\end{equation}
a renormalization constant $Z_m$ is needed, for the singlet combination
\begin{equation}
Z_m r_m\overline{m}
=Z_m r_m \frac{2m_{\ell}+m_s}{3}
=\frac{Z_m r_m}{6a}\left(\frac{2}{\kappa_{\ell}}+\frac{1}{\kappa_s}-
\frac{3}{\kappa_{\mathrm{crit}}}\right)=\frac{Z_A}{Z_P}
\overline{\widetilde{m}}
\end{equation}
the renormalization constant $Z_m \, r_m$ is introduced where $r_m>1$.
Note that $r_m$ depends on the gauge coupling and it turns out that in
the regime where the simulations are performed
$r_m$ can be rather large when determined non-perturbatively (up to $r_m\approx 2$).

The goal is now to determine the physical value of the
strange AWI quark mass $\widetilde{m}_s=\widetilde{m}_{s,\mathrm{ph}}$
as a function of $\kappa_\ell$ and  $\kappa_s \equiv \kappa_s(\kappa_\ell)$. This then will allow us to simulate at different pion masses, i.e.~different
values of $\kappa_\ell$, keeping $\widetilde{m}_s=\widetilde{m}_{s,\mathrm{ph}}$
constant. We start with the expression for the strange AWI mass
\begin{align}
3\widetilde{m}_s=2\left(\widetilde{m}_s-\widetilde{m}_{\ell}\right)+
3\overline{\widetilde{m}}
=\frac{Z}{2a}\left[2\left(\frac{1}{\kappa_s}-\frac{1}{\kappa_{\ell}}\right)
+r_m\left(\frac{1}{\kappa_s}+\frac{2}{\kappa_{\ell}}-\frac{3}{\kappa_{\mathrm{crit}}}\right)\right]\,,\label{eq:funny}
\end{align}
where $Z=Z_mZ_P/Z_A$. Setting $\widetilde{m}_s=\widetilde{m}_{s,\mathrm{ph}}$ gives
\begin{equation}
\frac{1}{\kappa_s} =  \frac{2}{2+r_m}
\left(\frac{3a}{Z}\widetilde{m}_{s,\mathrm{ph}} +
(1-r_m)\frac{1}{\kappa_{\ell}} + \frac{3 r_m}{2} 
\frac{1}{\kappa_{\mathrm{crit}}} \right).
\end{equation}
Subtracting the physical point result from both sides of the equation gives
\begin{equation}
 \frac{1}{\kappa_s} =\frac{1}{\kappa_{s,\mathrm{ph}}} +
\frac{2(1-r_m)}{2+r_m} \left(  \frac{1}{\kappa_{\ell}} - \frac{1}{\kappa_{\ell,\mathrm{ph}}} \right)\,,
\end{equation}
while the target $\kappa_{\ell}$ that corresponds to a given
$\widetilde{m}_{\ell}$ value can be obtained through
\begin{equation}
\frac{1}{\kappa_{\ell}}=\frac{1}{\kappa_{\ell,\mathrm{ph}}}+
\frac{2a(2+r_m)}{3Zr_m}(\widetilde{m}_{\ell}-\widetilde{m}_{\ell,\mathrm{ph}}).
\end{equation}
The combination $Z$ is extracted by fitting the AWI masses along the $\overline{m}=\text{const.}$ line as a function of $\kappa_\ell$
and $\kappa_s$. Along the same trajectory we define the physical point as
the point where
$\widetilde{m}_{s}/\widetilde{m}_{\ell}$ assumes its physical value
of 27.46(44)~\cite{Aoki:2016frl}. This gives $\kappa_{\ell,{\mathrm{ph}}}$ and $\kappa_{s,{\mathrm{ph}}}$.
Finally, $Zr_m$ (and $\kappa_{\mathrm{crit}}$ if needed) can be obtained
from $\widetilde{m}$ as a function of $1/\kappa$ along the symmetric $m_s=m_{\ell}$ line.

Note that when full $\mathcal{O}(a)$ improvement is carried out,
four additional combinations of improvement coefficients
appear. These can be fitted in a similar fashion without
additional effort, the formulae however become more involved,
see Ref.~\cite{Bali:2016umi}.

\begin{figure}[t]
\centerline{\includegraphics[width = 0.48\textwidth]{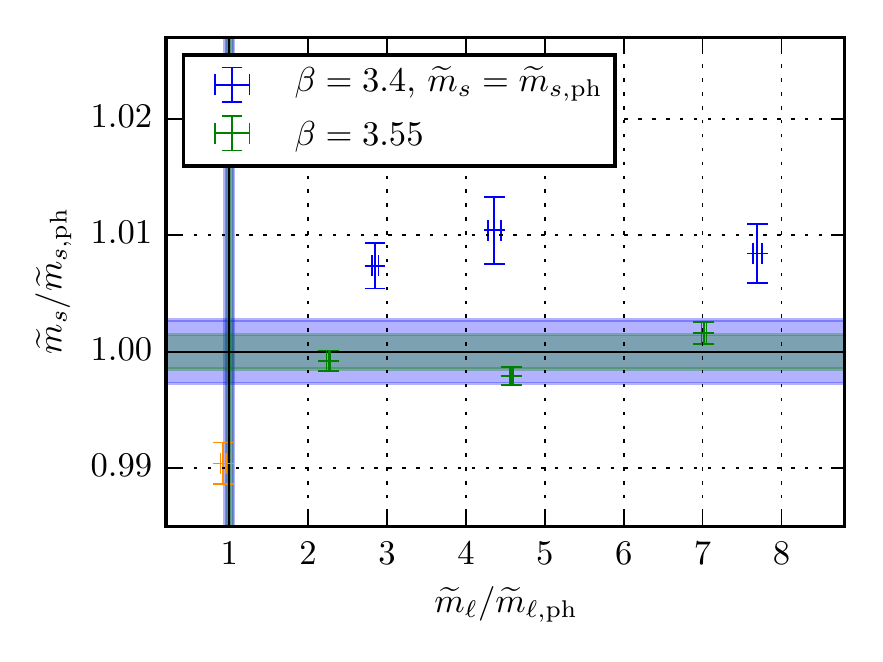}
    \includegraphics[width = .48\textwidth]{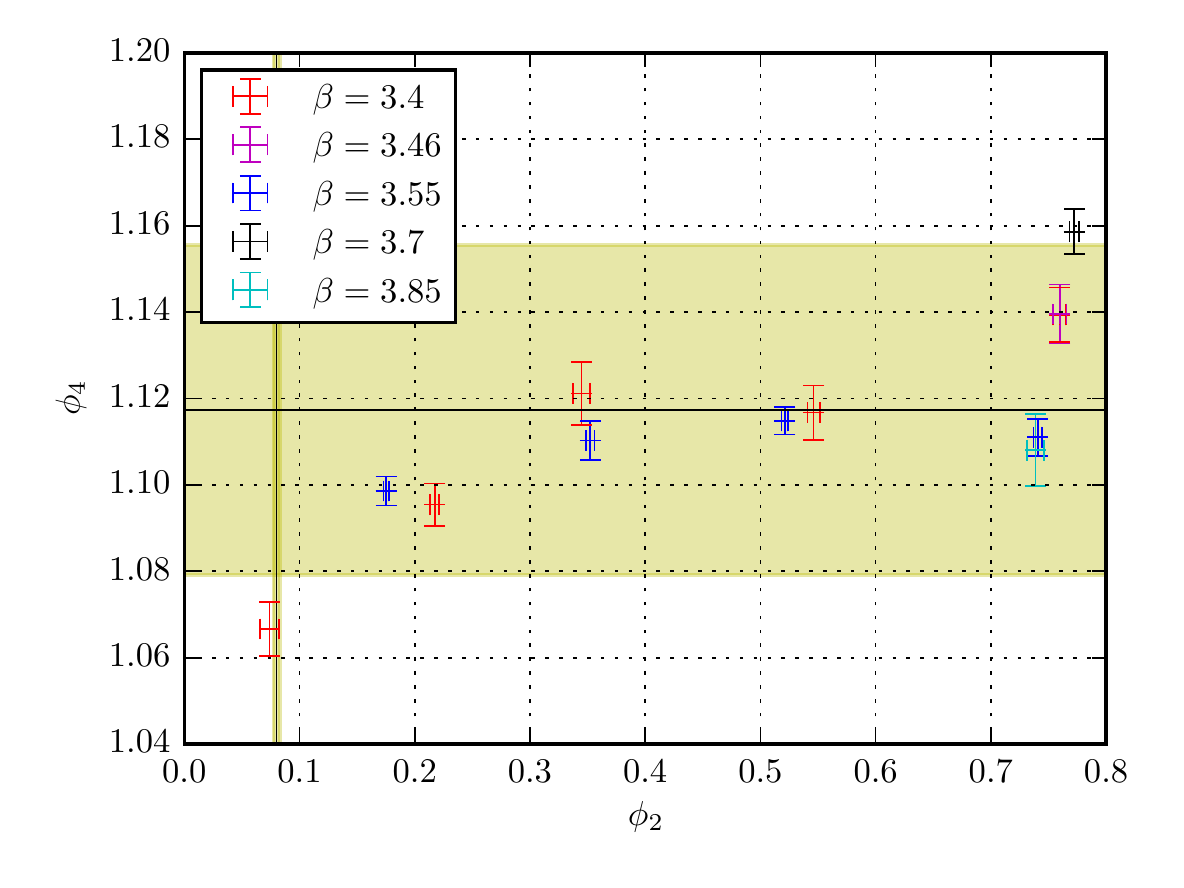}}
    \centerline{\includegraphics[width = .48\textwidth]{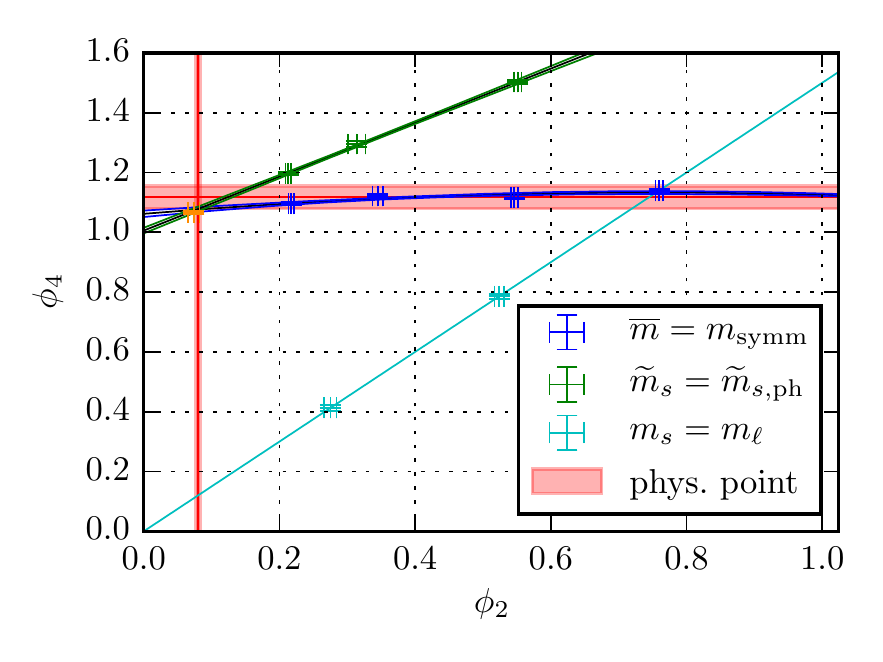}
    \includegraphics[width = .48\textwidth]{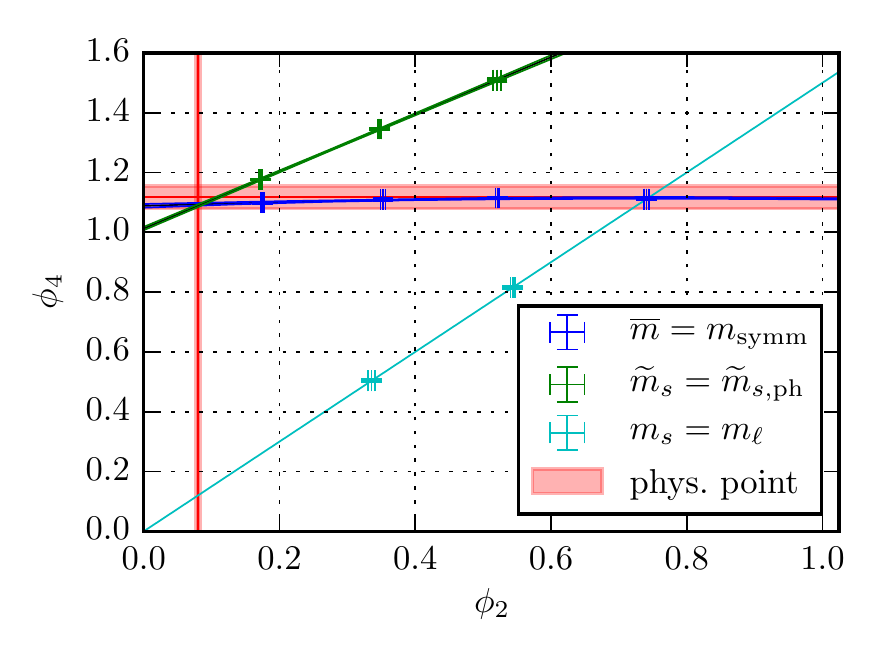}}
\caption{\label{fig:tuning} Top left: $\widetilde{m}_s$ vs.~$\widetilde{m}_\ell$ along the chiral trajectory $\widetilde{m}_s =\widetilde{m}_{s,\mathrm{ph}}$.
Top right: $\phi_4$ vs.~$\phi_2$ along $\overline m = m_\mathrm{symm}$.
Lower plots: chiral extrapolations for $\beta=3.4$ and $\beta=3.55$ of $\phi_4$ vs.~$\phi_2$ for different chiral trajectories. Plots (except upper right plot) taken from Ref.~\cite{Bali:2016umi}.}    
\end{figure}

\begin{figure}[t]
    \centerline{\includegraphics[width = 0.48\textwidth]{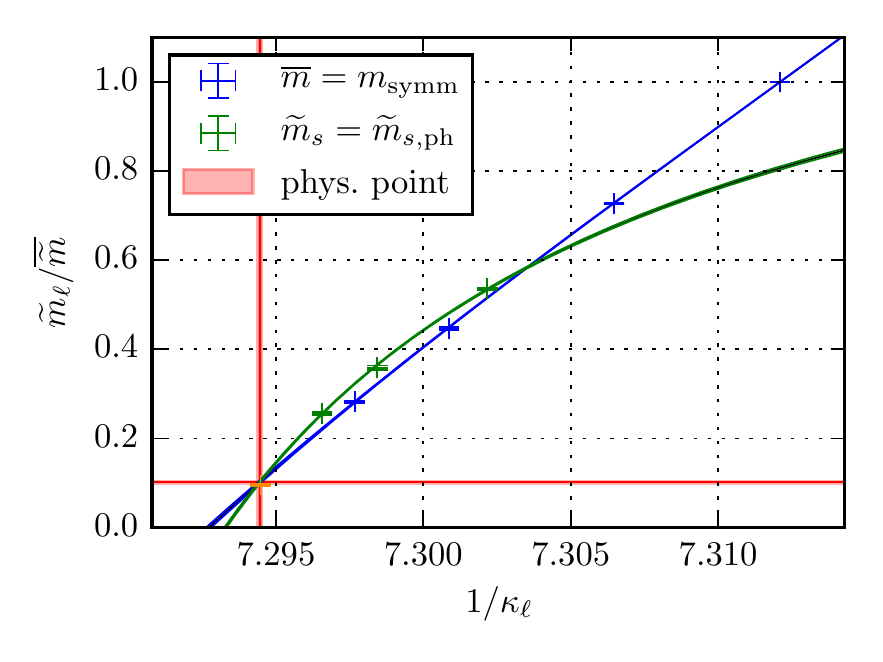}
    \includegraphics[width = 0.48\textwidth]{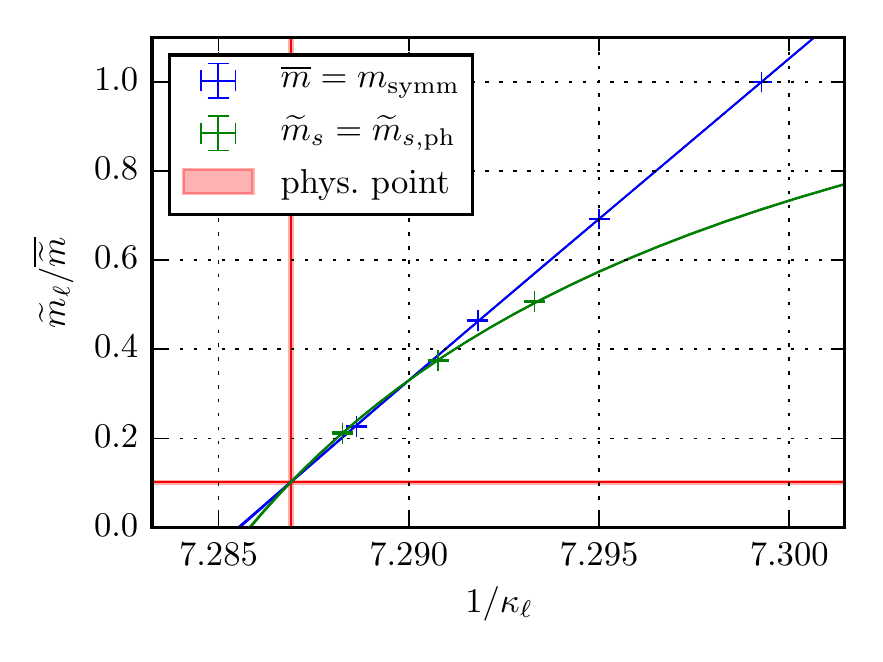}}
    \centerline{\includegraphics[width = 0.48\textwidth]{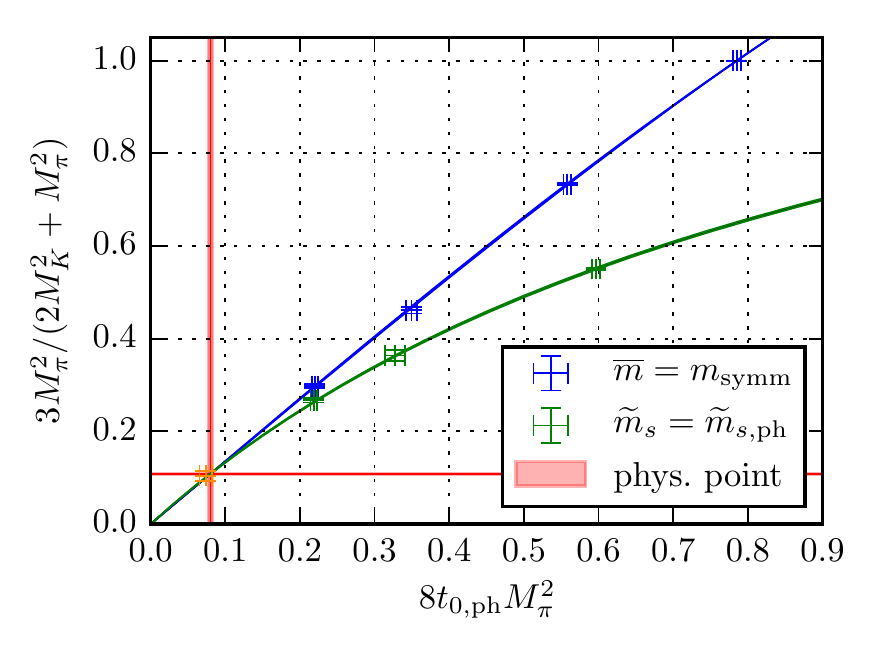}
    \includegraphics[width = 0.48\textwidth]{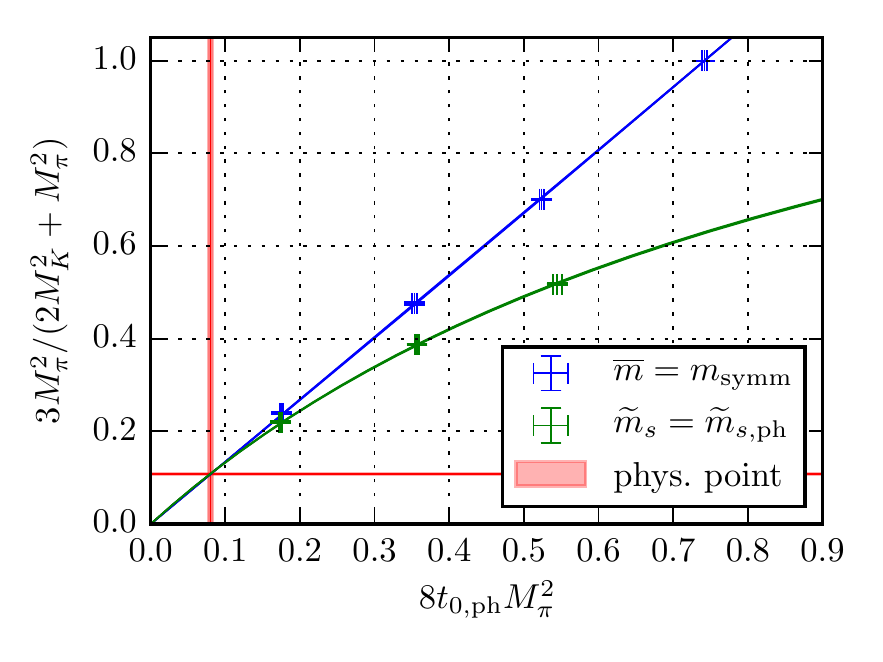}}
    \caption{\label{fig:phys}Determination of the physical point for $\beta=3.4$ (left) and  $\beta=3.55$ (right) using the ratio of quark masses (upper plots).
    We find good agreement with the corresponding ratio of meson masses (lower plots). Figures taken from Ref.~\cite{Bali:2016umi}.}
\end{figure}
\section{How well does the tuning of the simulation parameters work?}
Here we investigate the possibility, that one or both of the chiral trajectories, $\overline m = m_\mathrm{symm}$ and $\widetilde{m}_s =\widetilde{m}_{s,\mathrm{ph}}$,
could have been, in principle, mistuned.
We start with the $\widetilde{m}_s =\widetilde{m}_{s,\mathrm{ph}}$
trajectory. The most obvious question is how well
$\widetilde{m}_s$ is kept constant.
The quality of our tuning is visualized in the top left plot
of Fig.~\ref{fig:tuning}. 
The vertical and horizontal bands
correspond to propagated errors
from the determination of the physical light and strange AWI quark
masses as outlined in the previous section.
For $\beta=3.55$ the data points
match the prediction within per mille accuracy. For $\beta=3.4$ we
find that $\widetilde{m}_s$ also remains constant within tiny errors,
however, there appears to be a 1\% shift
relative to the prediction. The origin of this can be traced back
to an updated value of the improvement coefficient $c_A$ that
only became available when the simulations had already been started.
The $\kappa$ values at which the simulations have been performed come from a prediction based on the previous estimate of $c_A$
while the prediction shown in the figure (blue band) is based on the new value of $c_A$.
In any case, such a small misalignment is not of practical relevance.

Another reason for concern is whether the $\overline{m} = m_{\mathrm{symm}}$
trajectory hits the physical point. If this were not the case then also
the $\widetilde{m}_s =\widetilde{m}_{s,\mathrm{ph}}$ trajectory
would have to be somewhat reweighted. 
In Fig.~\ref{fig:tuning} (top right) we plot $\phi_4 = 8 t_0 (m_K^2 + m_\pi^2/2) \sim \overline m$ versus $\phi_2=t_0 m_\pi^2 \sim m_l$ for the
$\overline m = m_\mathrm{symm}$ ensembles across different
lattice spacings. The yellow horizontal and (barely visible) vertical
bands correspond to the physical point values
within their present uncertainties. One aim of our simulations is to ultimately
reduce these errors. There is no indication that, extrapolated to
the physical $\phi_2$ value, our simulation points are at variance with the
target range. At $\beta=3.4$, however, we observe a significant slope
which becomes negligible towards the finer lattice spacings.

\begin{figure}[t]
\centerline{\includegraphics[width = .48\textwidth]{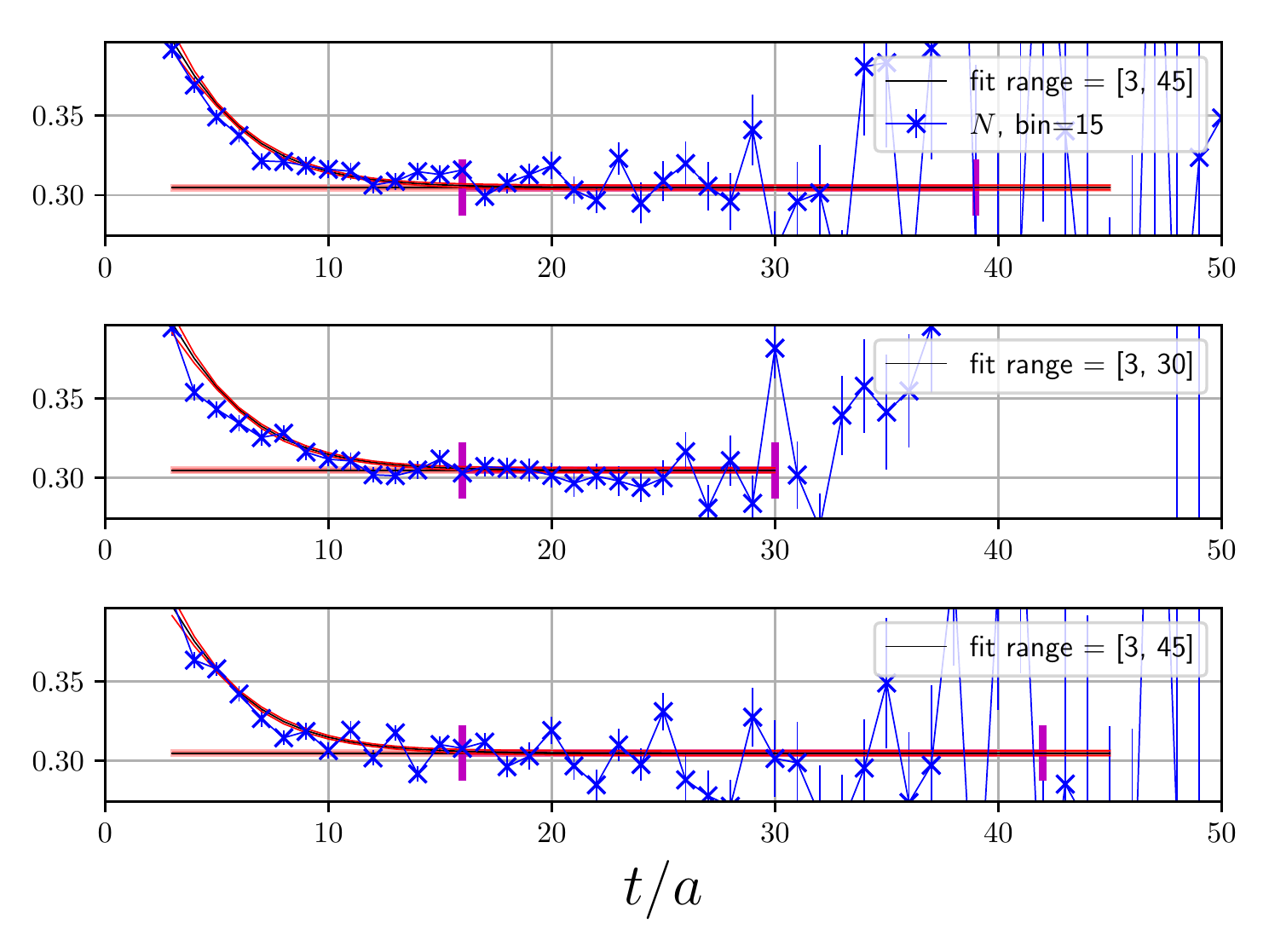}
\includegraphics[width = .48\textwidth]{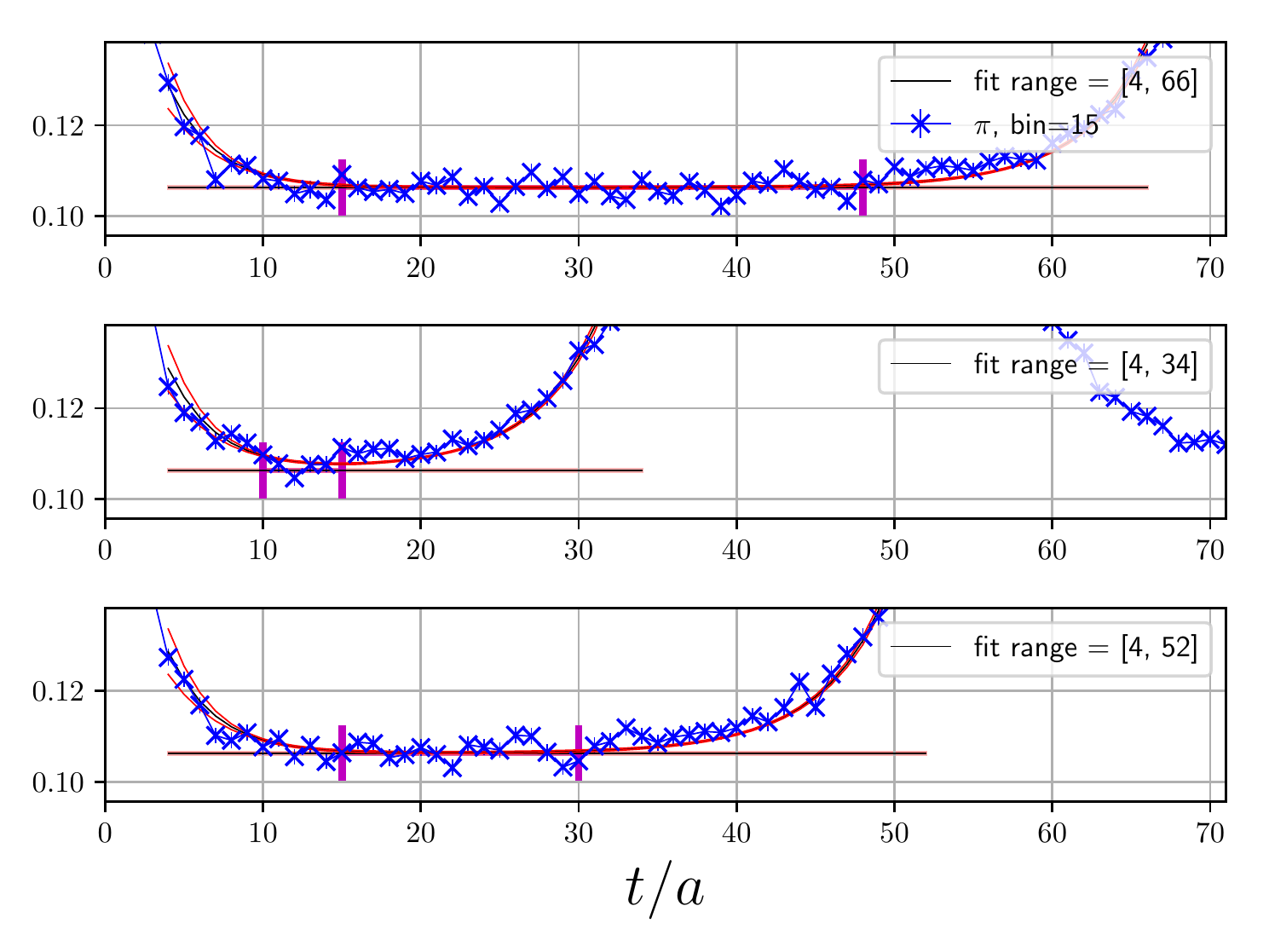}}
\caption{\label{fig:effmass}Effective mass for the nucleon (left) and pion (right) on the N300 ensemble.
The temporal source positions of the point-to-all correlators are
$t_{\rm src}/a=46,81,63$ (top to bottom), $t/a$ corresponds to the distance between source and sink in forward-direction. The lattice has the time extent $T=127a$.
The vertical red bands indicate the determined fit range by the procedure described in the text.}
\end{figure}

At $\beta=3.4$ and $\beta=3.55$ we have investigated the chiral
extrapolations in more detail, see the lower plots of Fig.~\ref{fig:tuning}.
Note that $\phi_4$ is constant to next-to-leading order
chiral perturbation theory (NLO $\chi$PT) along $\overline{m}=\text{const}$.
Corrections are of higher order in the quark mass or due to
discretization effects.
The dependence on $\phi_2$ is weaker at the smaller lattice spacing
($\beta=3.55$) which may indicate this is a lattice artefact.
Most importantly, at the physical point we are within the target range.

We remark that the orange data points in Fig.~\ref{fig:tuning} (and all other figures) correspond to the D100 ensemble~\cite{Bali:2016umi} (see Fig.~\ref{fig:simdetail})
that has only extremely limited
statistics and therefore in this case the errors should be regarded with caution. Hence, this ensemble was excluded from any further analysis. Nevertheless,
the data at this stage do not suggest any major surprises.

We remind the reader that the physical point was determined on the
$\overline m = m_{\mathrm{symm}}$ line.
This determination is shown in the upper panels of Fig.~\ref{fig:phys}
for $\beta=3.4$ and $\beta=3.55$, where we plot
$\widetilde{m}_{s}/\widetilde{m}_{\ell}$ versus $1/\kappa_\ell$.
The lines correspond to a global fit of AWI quark masses including 
$\mathcal O(a)$ improvement. The value of $\kappa_{\ell}$ where
$\widetilde{m}_{s}/\widetilde{m}_{\ell}  = 27.46(44)$~\cite{Aoki:2016frl}
then defines the physical point. Note that the green curves are predictions,
whereas the corresponding data points are the results
of subsequent simulations.

The FLAG~\cite{Aoki:2016frl} value of the ratio of quark masses that we
used relies on input from lattice simulations.
As a cross-check we compare our data against the
corresponding meson mass ratio $3M_{\pi}^2/(2M_K^2+M_{\pi}^2)$:
the lower plots of Fig.~\ref{fig:phys} demonstrate excellent
agreement of the meson mass ratio with the physical point value
(red lines). Our errors mean that we will be able to improve the
precision of the ratio of quark masses relative to the value
quoted in the FLAG report,
once a continuum limit extrapolation has been carried out.

\begin{figure}[t]
\centerline{
    \includegraphics[width = 0.45\textwidth]{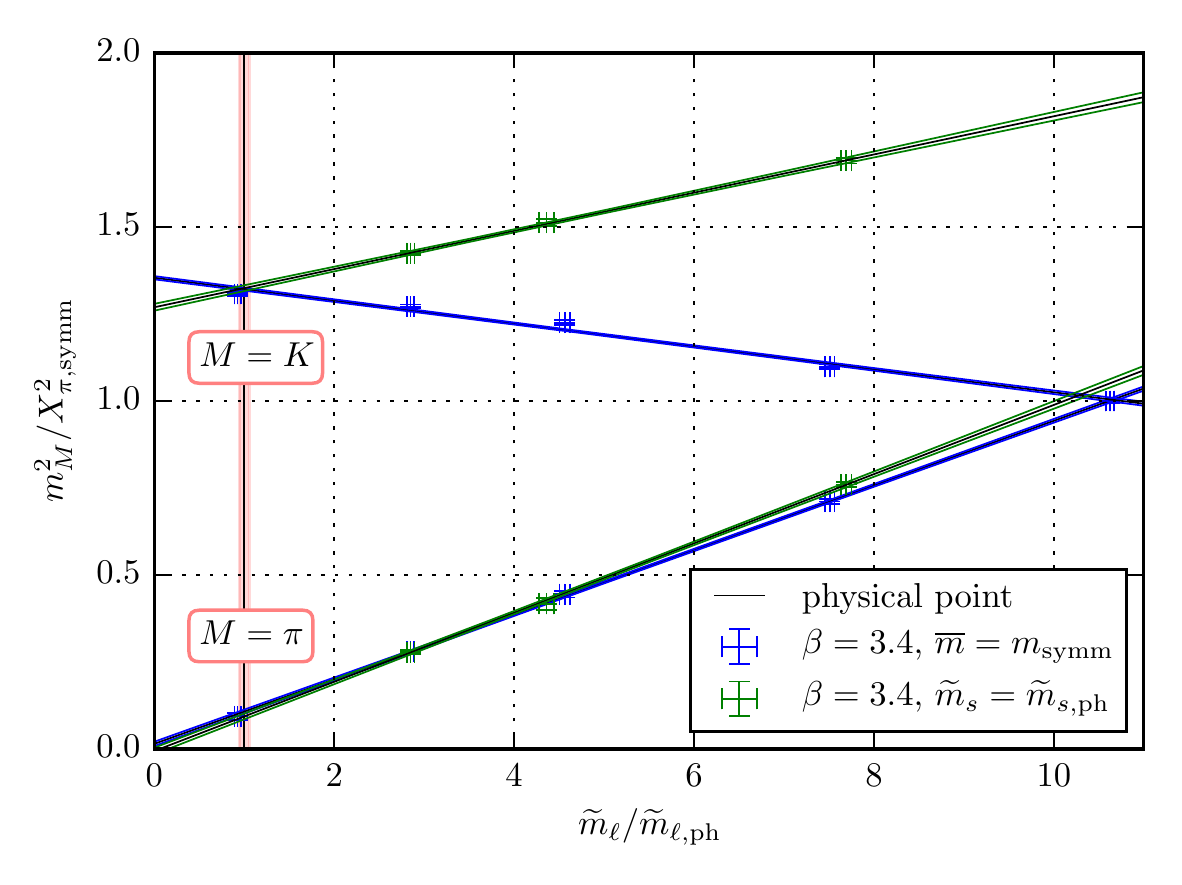}
    \includegraphics[width = 0.45\textwidth]{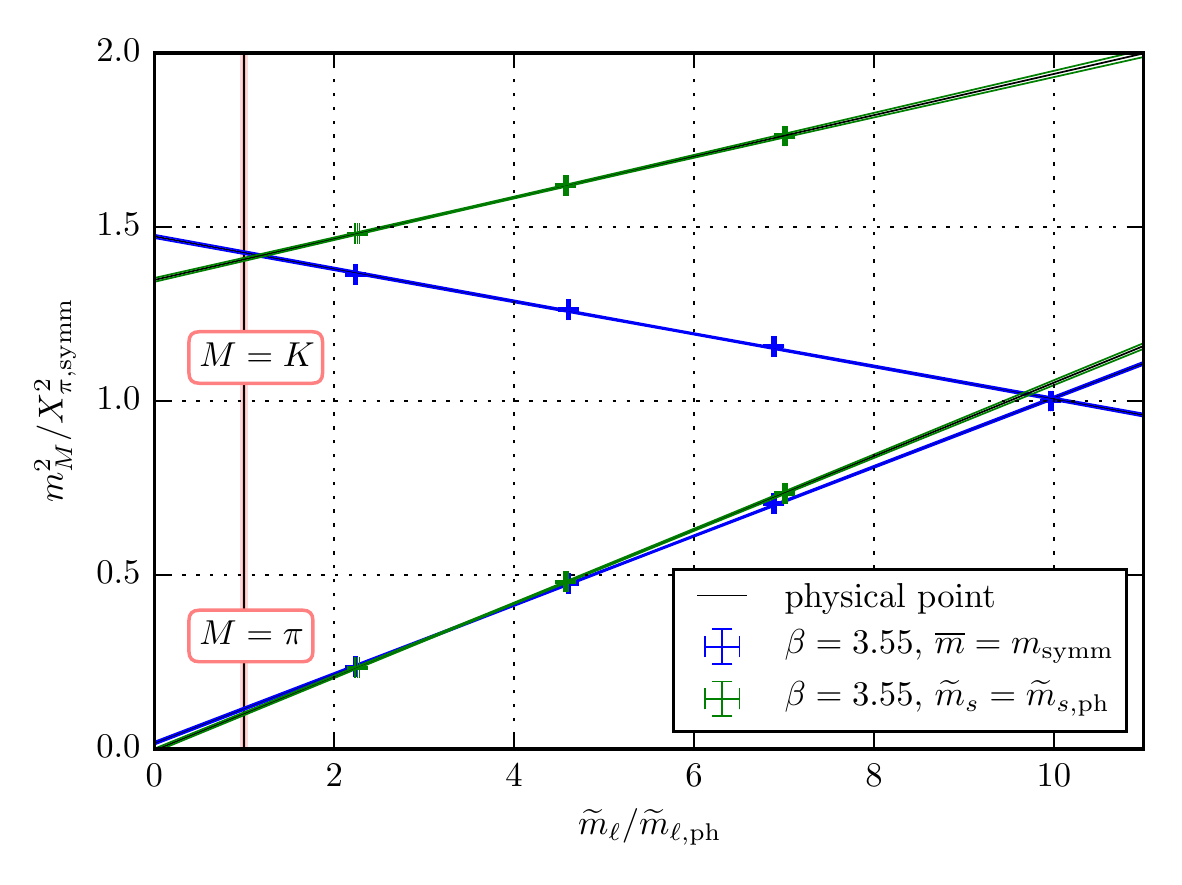}}
\centerline{
    \includegraphics[width = 0.45\textwidth]{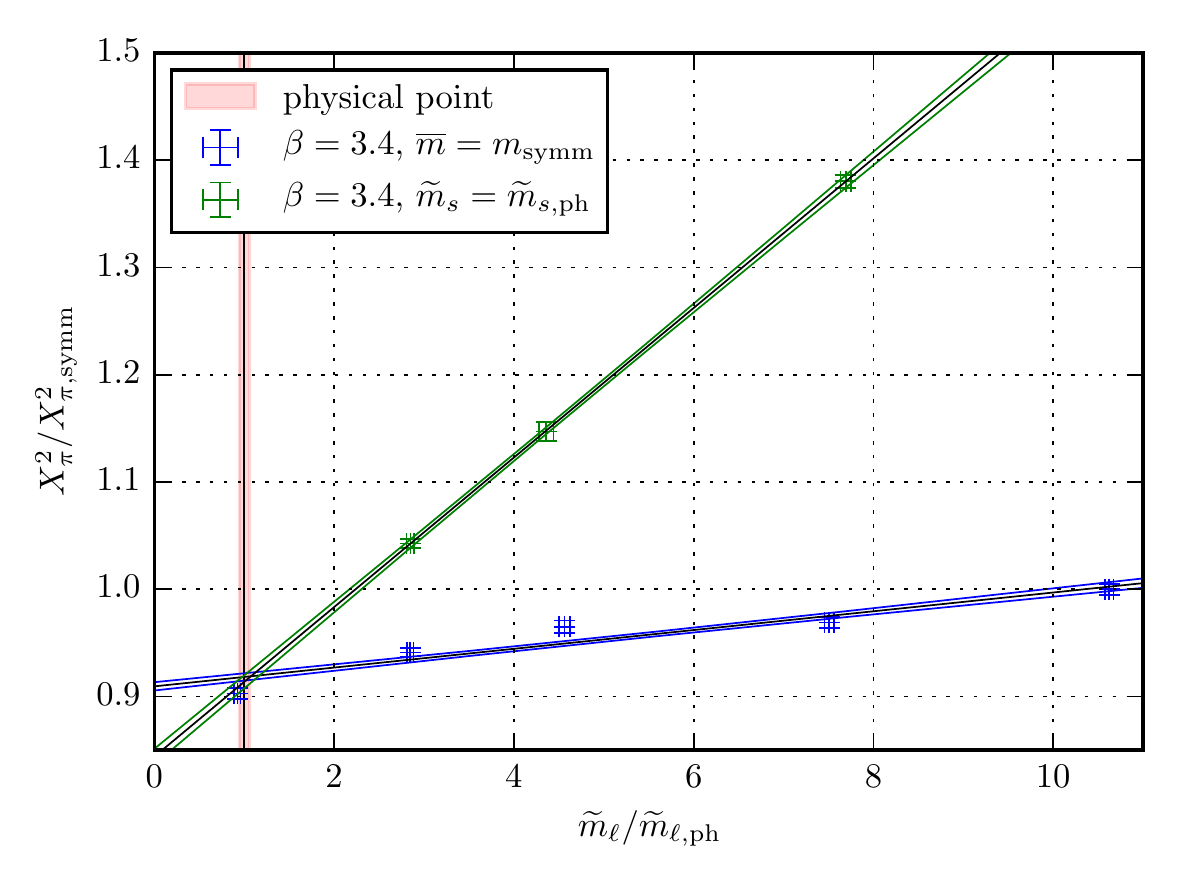}
    \includegraphics[width = 0.45\textwidth]{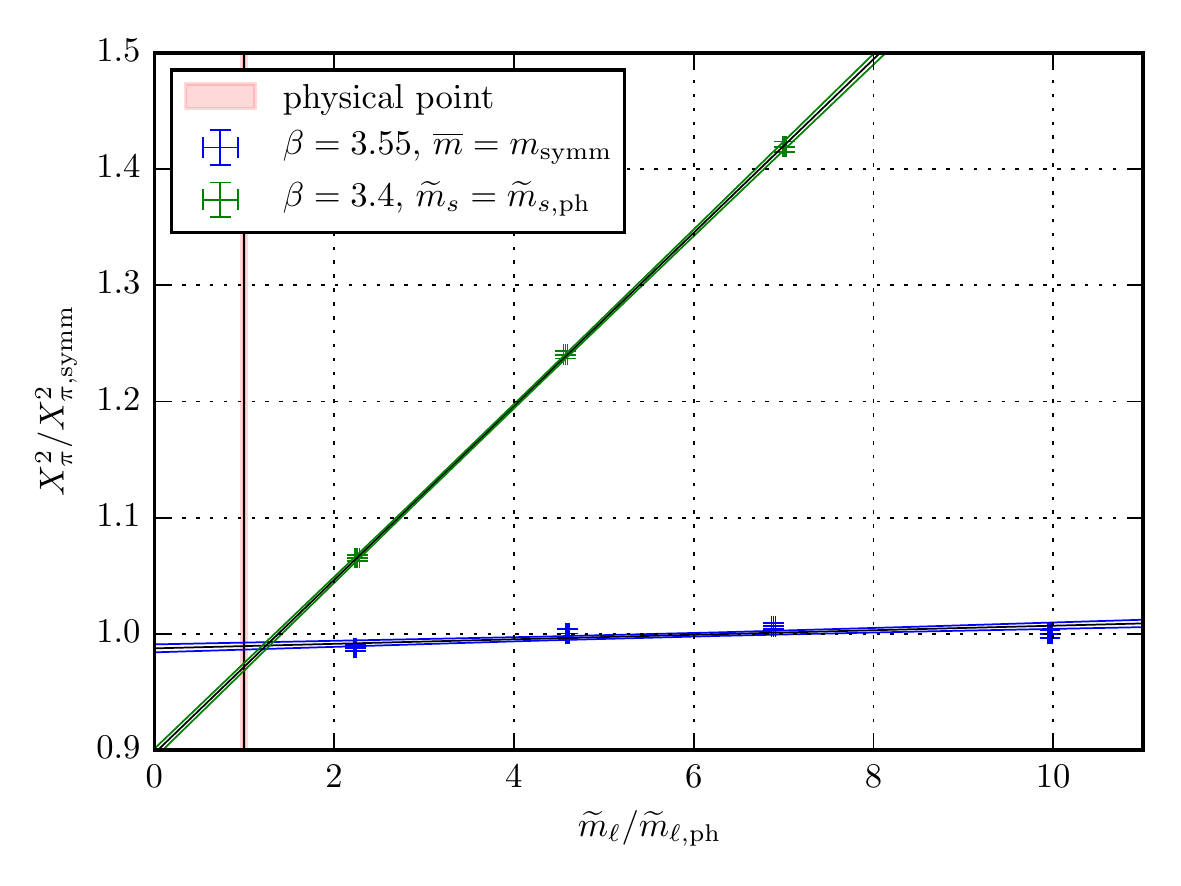}}
\caption{\label{fig:ps}Pion and kaon masses (upper plots) and average meson masses (lower plots), normalized to the average meson mass
of the symmetric point for $\beta=3.4$ (left) and $\beta=3.55$ (right) along
$\overline m = m_\mathrm{symm}$ (blue) and $\widetilde{m}_s=\widetilde{m}_{s,\mathrm{ph}}$ (green).
The vertical red bands correspond to the physical values.}
\end{figure}

    \section{Measurement Details\label{sec:meas}}
The light meson and baryon spectra are obtained from smeared-smeared point-to-all correlators where the sources are placed deep within the bulk
of the lattice to avoid effects from the open boundaries in time. The
spatial source positions were chosen randomly.
For the computation of the pion and kaon masses we additionally make use of the one-end trick.
The temporal source positions for these point-smeared and point-point
correlators are always placed close to a boundary.

To compute the baryon spectrum we use standard relativistic interpolators,
e.g., we destroy the nucleon applying $N_{\alpha} = \epsilon_{ijk} u^i_{\alpha} \left( u^{jT} C \gamma_5 d^k\right)$, where for the quark fields
we use Wuppertal smearing on 3-dimensionally APE smeared gauge links.
For the computation of the correlators a custom version of the {\sc CHROMA}
software package~\cite{Edwards:2004sx} including the {\sc LibHadronAnalysis}
library
has been developed where also the multigrid solver implementation
of Refs.~\cite{Heybrock:2015kpy,Frommer:2013fsa} is used.

The fitting procedure is based on a two stage process. We first determine the starting point of the actual fit range by means of a double exponential fit;
the time slice where the contribution of the excited
state becomes negligible defines the beginning of the fit range.
The end in the case of baryons is determined as the time slice where the signal to noise ratio of the correlator becomes small.
For pseudoscalar mesons it is a time slice where the contribution
to the correlator of states
propagating from the opposite boundary is negligible.
This is estimated by fitting the correlator to the functional form of a
hyperbolic sine close to the boundary.
For illustration of this first step we show the effective masses, fits
and fit ranges for the nucleon
and the pion correlators
for the N300 ensemble ($a\approx 0.05\,\text{fm}$) in Fig.~\ref{fig:effmass}.
In a second step we then extract the masses within the determined fit range by fitting to a single exponential.
Autocorrelations are taken into account by means of a binning analysis
where we extrapolate the error to infinite bin size.

\begin{figure}[t]
\centerline{
     \includegraphics[width = 0.45\textwidth]{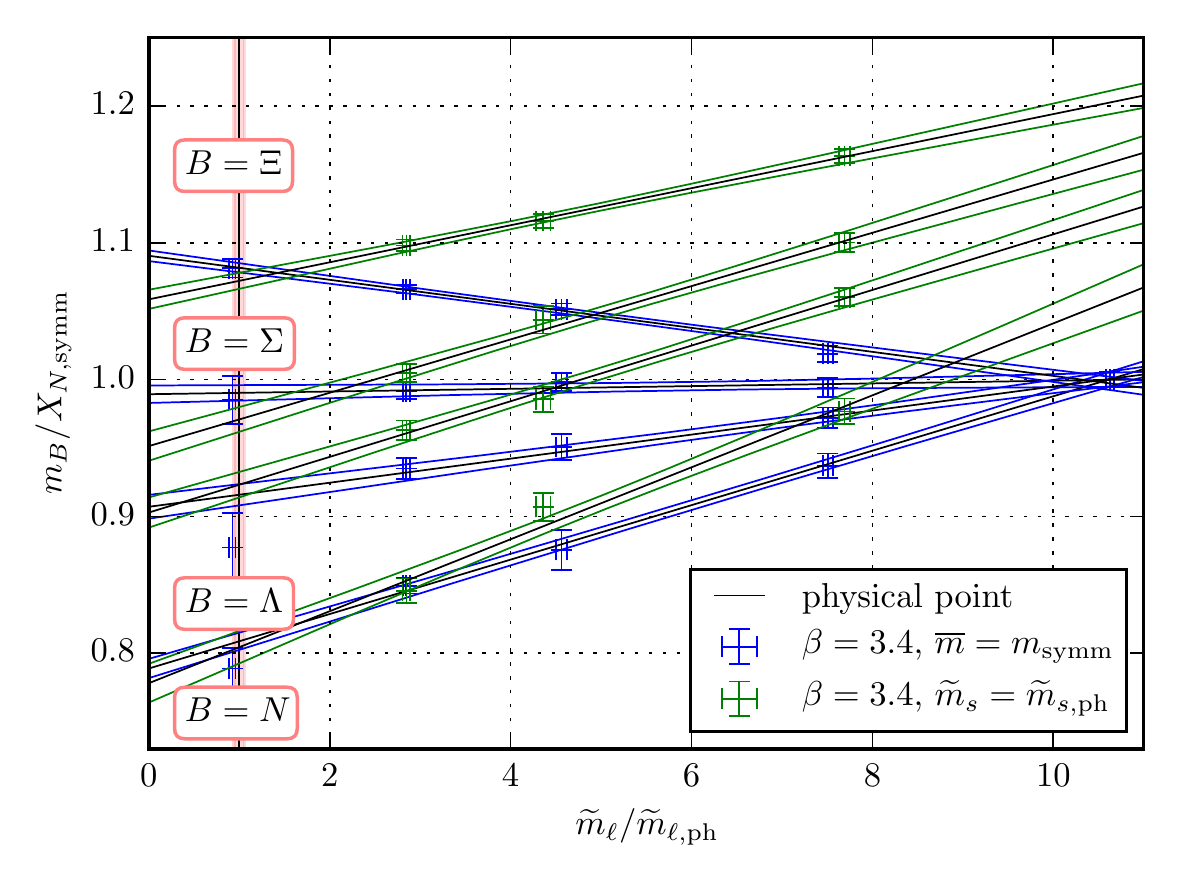}
    \includegraphics[width = 0.45\textwidth]{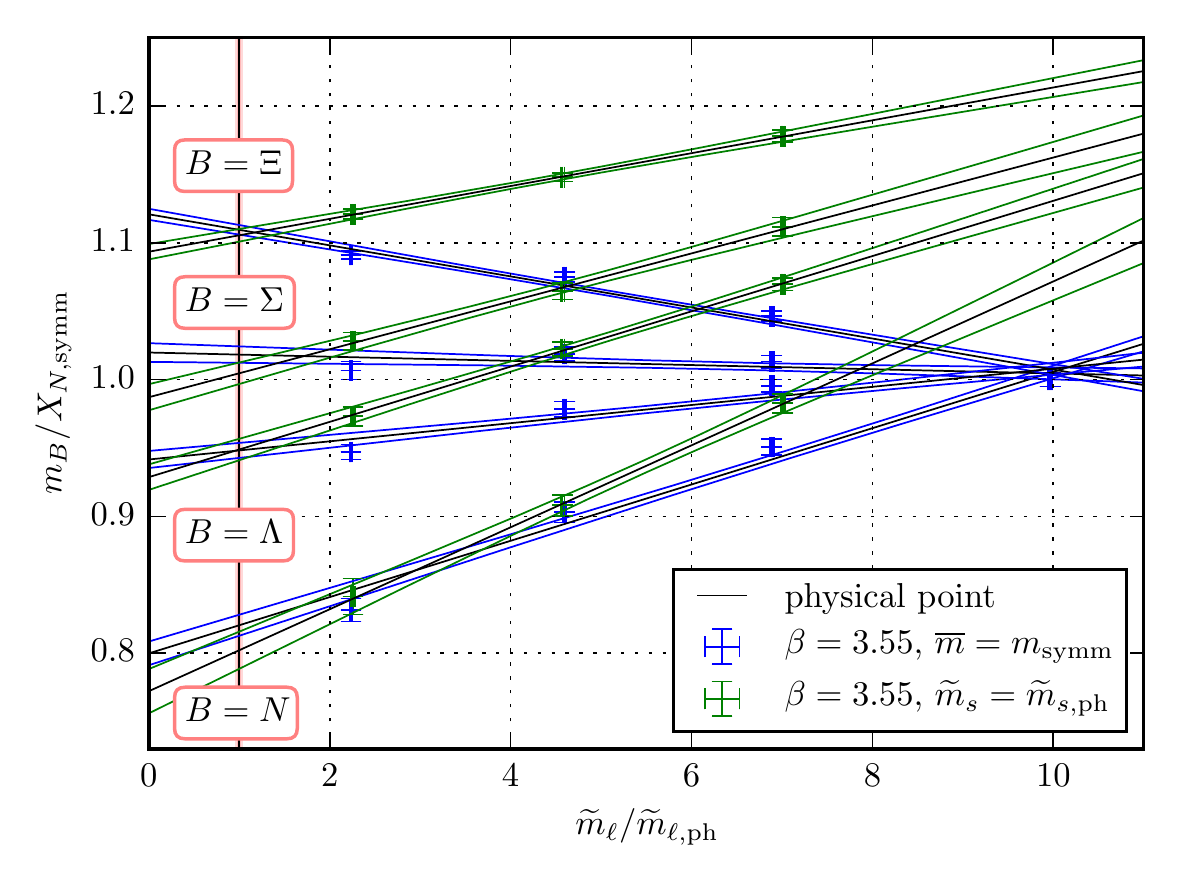}}
\centerline{
    \includegraphics[width = 0.45\textwidth]{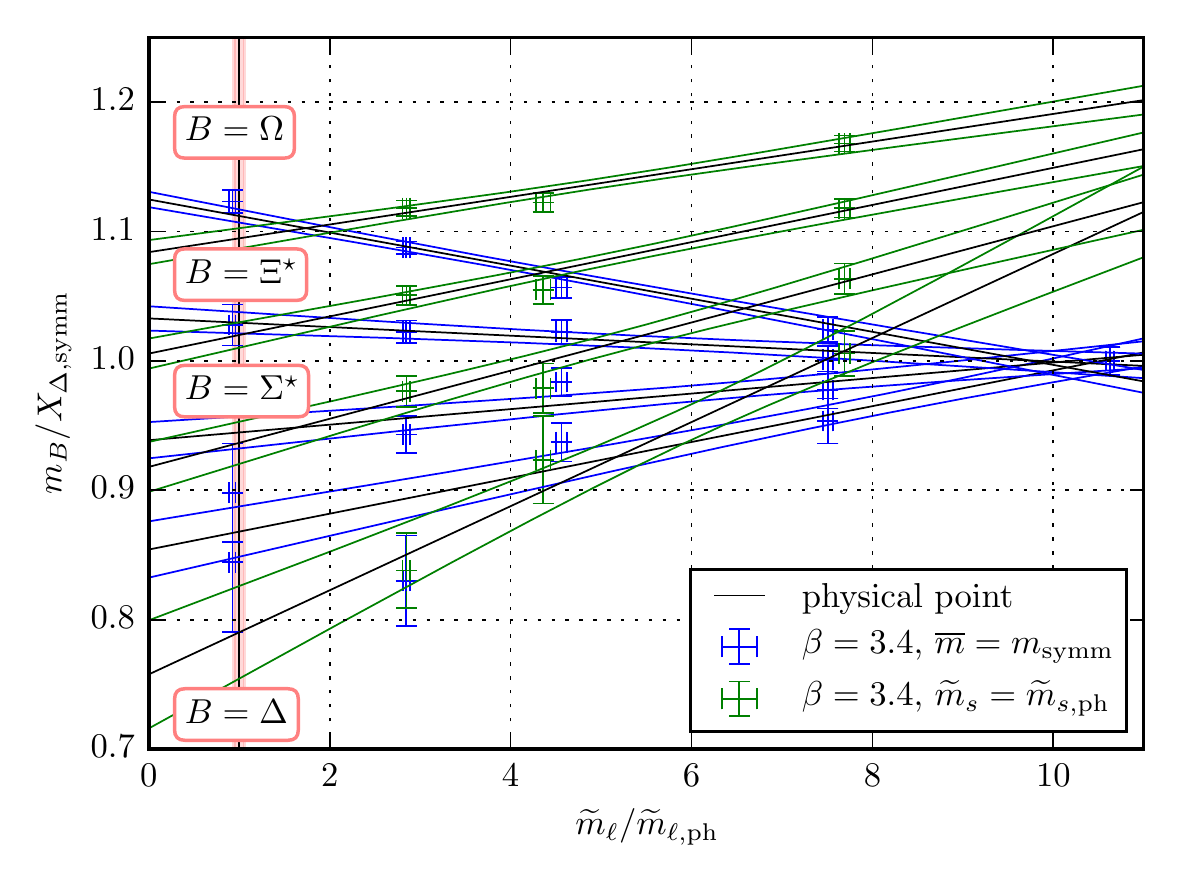}
    \includegraphics[width = 0.45\textwidth]{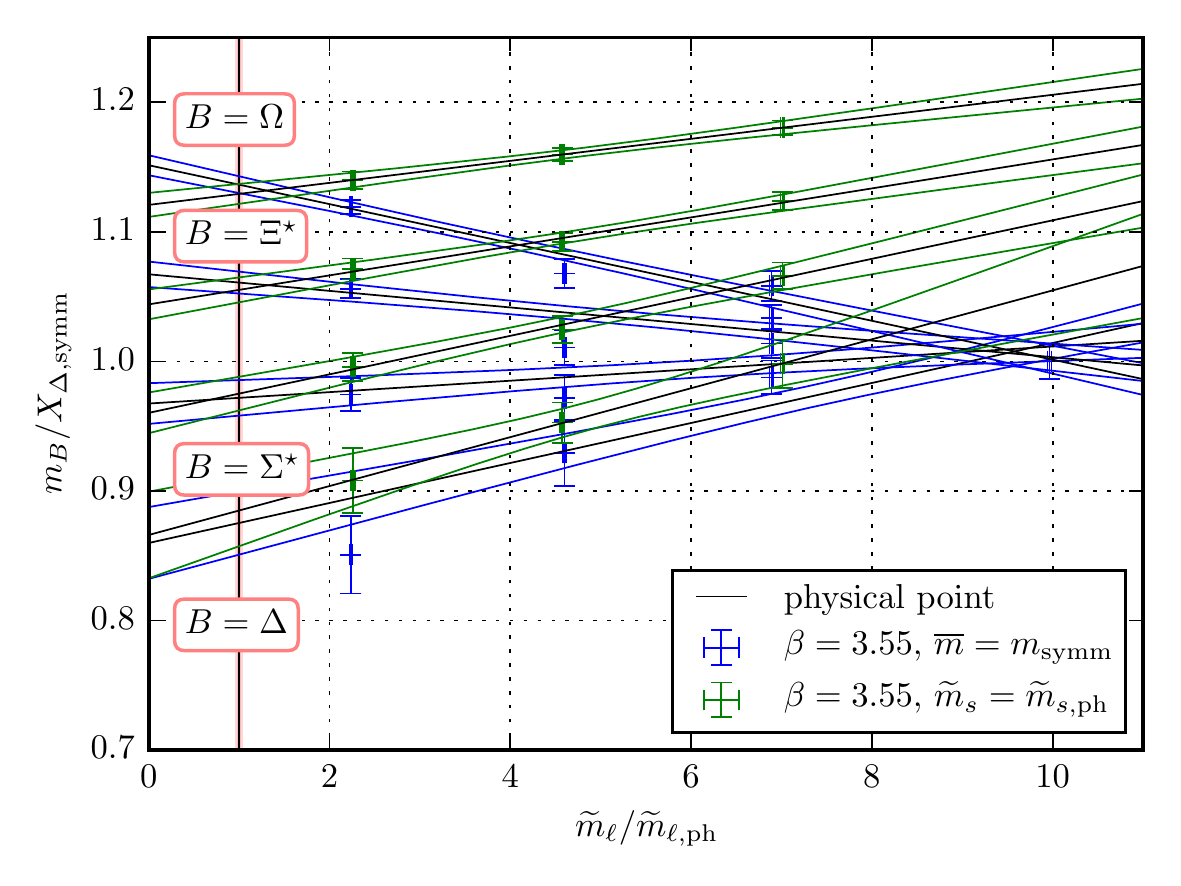}}
\caption{\label{fig:oct}Baryon octet masses (upper plots) and decuplet masses (lower plots) normalized by the average octet, respectively decuplet, mass
  of the symmetric point for $\beta=3.4$ (left) and $\beta=3.55$ (right) along the two chiral trajectories
  $\overline m = m_\mathrm{symm}$ (blue) and $\widetilde{m}_s=\widetilde{m}_{s,\mathrm{ph}}$ (green).
  The vertical red bands correspond to the physical values.}
\end{figure}

\section{Comparing chiral extrapolations along $\overline m = m_\mathrm{symm}$ with $\widetilde{m}_s=\widetilde{m}_{s,\mathrm{ph}}$}
\label{sec:baryon}
As outlined in Sec.~\ref{sec:sim} we have two chiral trajectories at hand for extrapolating to the physical point. It is of particular interest to
check how consistent the results will be for masses not directly related
to the quark mass ratio used to define the physical point.

We start with the pion and kaon masses at the $\beta$ values where
both chiral trajectories are available, $\beta=3.4$ and $\beta=3.55$.
The squared average meson mass is defined as
$X_\pi^2 \equiv \left( 2 m_K^2 + m_\pi^2  \right)/3$.
Note that $\phi_4$, which has been used for the tuning process, is proportional to $X_\pi^2$: $\phi_4 = 12 t_0 \,X_\pi^2$. Also $X_\pi^2$, as well as $t_0$, is
constant along the $\overline m = m_\mathrm{symm}$ line in 
NLO $\chi$PT. In the upper plots of Fig.~\ref{fig:ps} we show the pion and kaon masses
along both trajectories, normalized with respect to the average meson mass determined at the symmetric point, as functions of the ratio of the light AWI quark
mass over its physical point value.
In this way we can compare results obtained at different values of $\beta$ in a way that is independent of the lattice spacing and renormalization constants.
The lines shown correspond to linear fits to the respective
pairs of data sets, without constraining them to coincide at the physical
point. We find the kaon mass lines intersect nicely at the physical point.
This also holds for the average meson mass, see the lower plots in
Fig.~\ref{fig:ps}. Only in one case we find a small
difference between the intersection point and the physical point.
For the moment being we aimed for a qualitative comparison and neglected
effects due to the lattice spacing and/or higher orders of $\chi$PT.
A detailed analysis is currently ongoing.

\begin{figure}[t]
\centerline{
    \includegraphics[width = 0.45\textwidth]{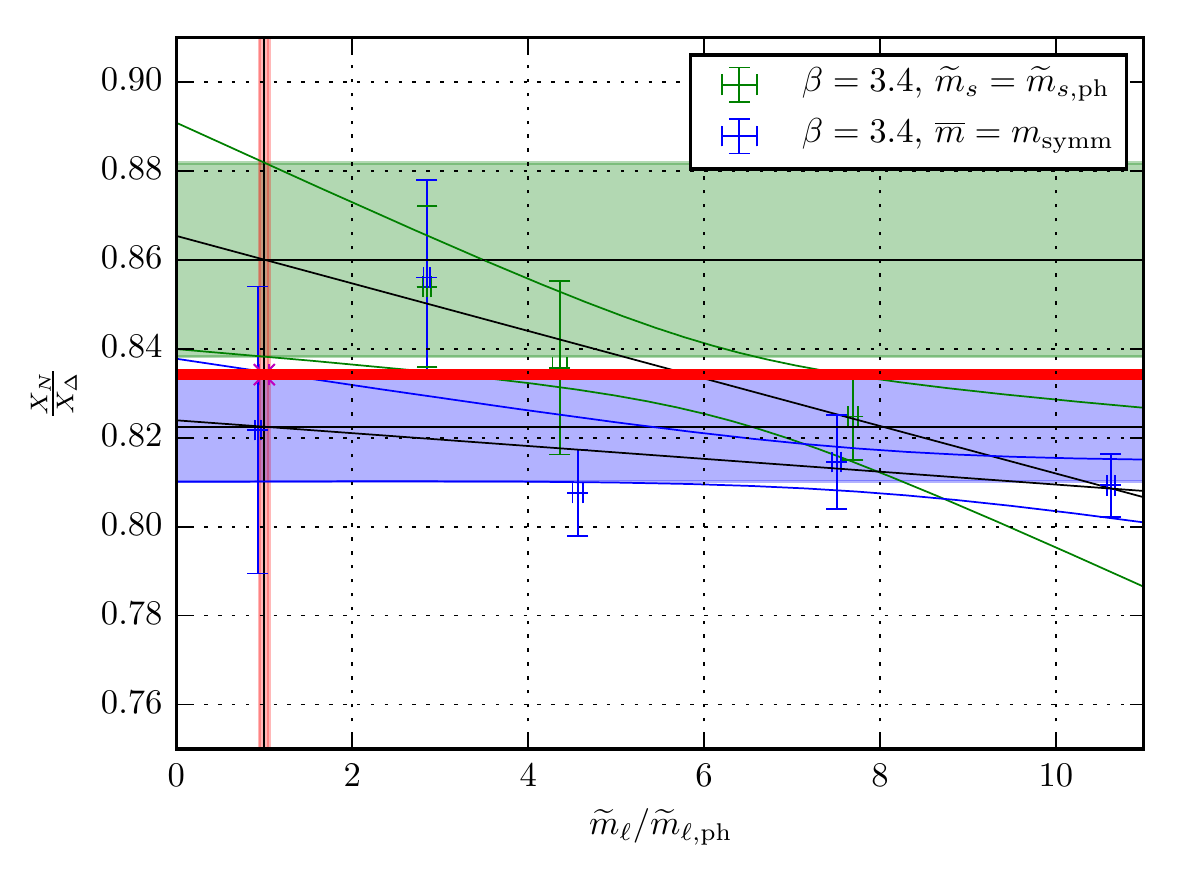}
    \includegraphics[width = 0.45\textwidth]{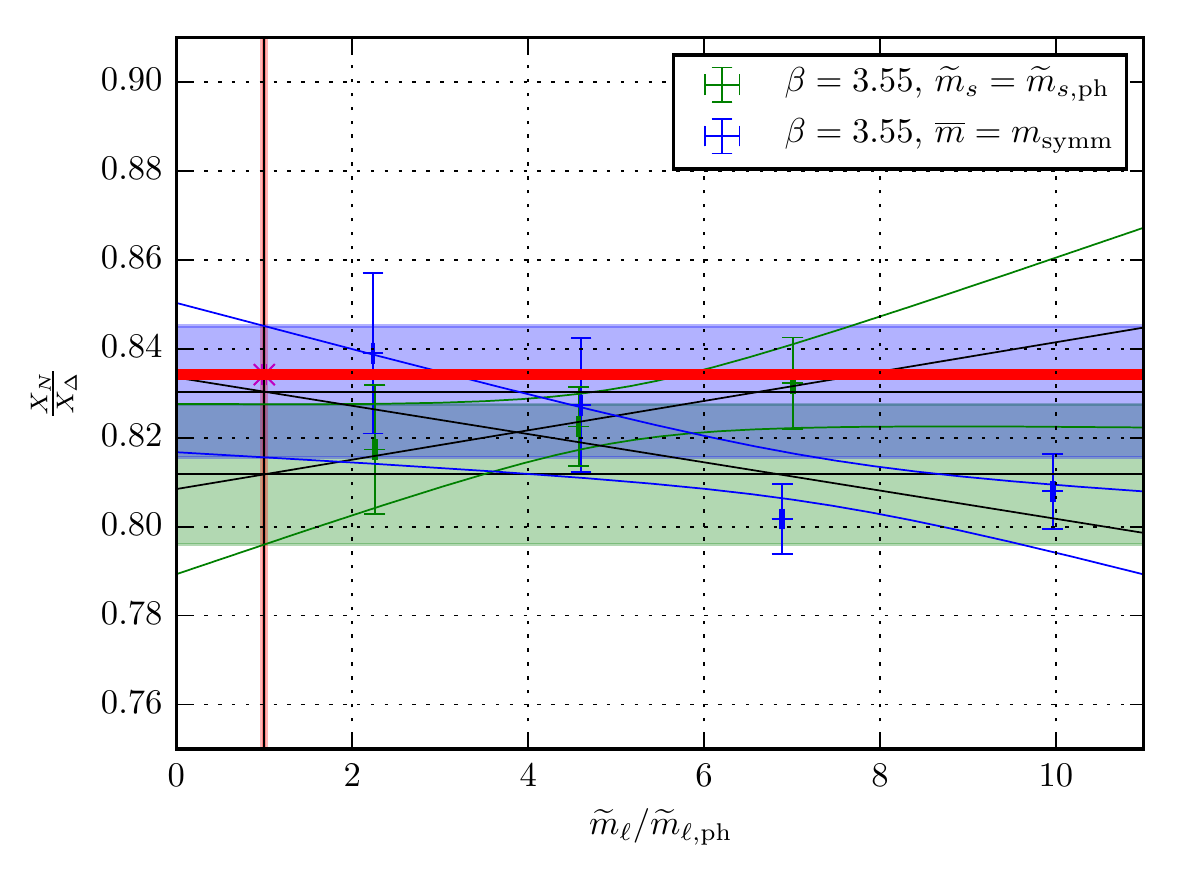}}
\centerline{     
     \includegraphics[width = 0.45\textwidth]{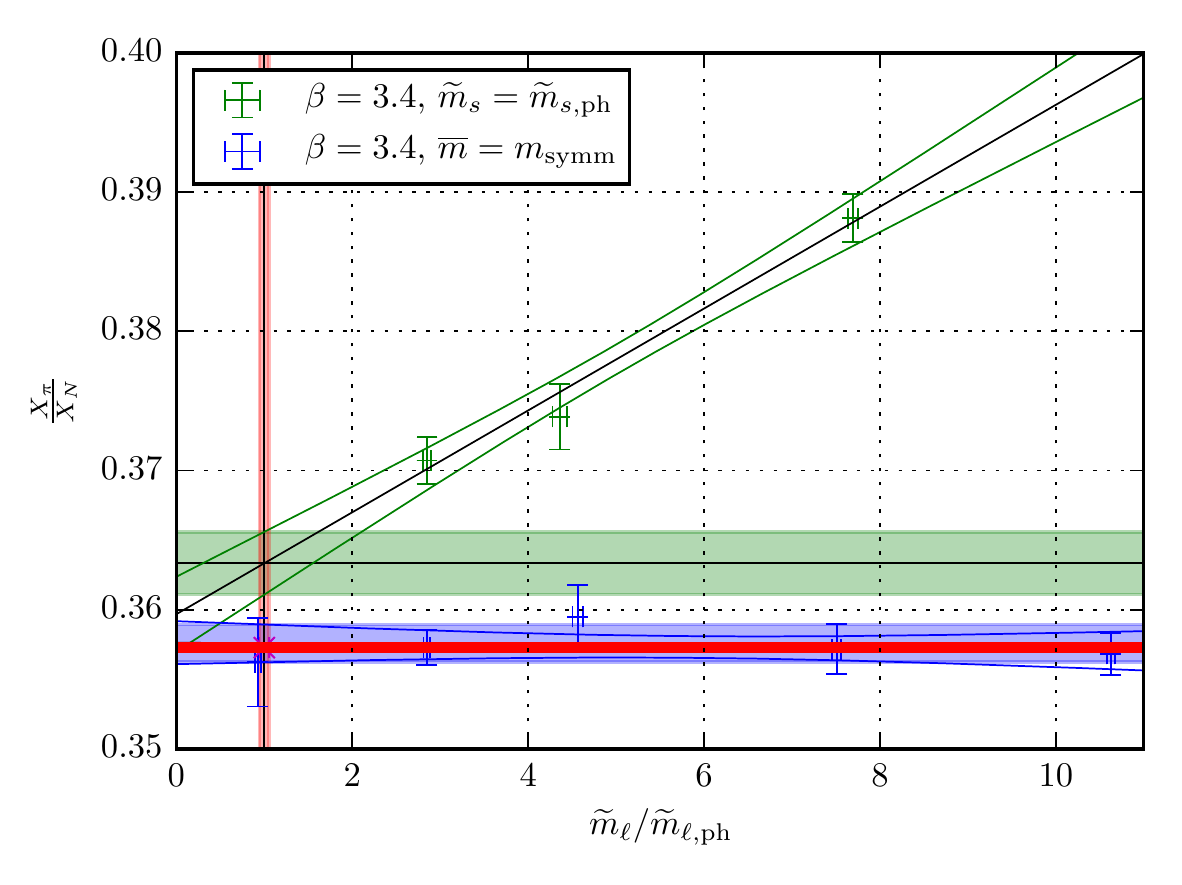}
     \includegraphics[width = 0.45\textwidth]{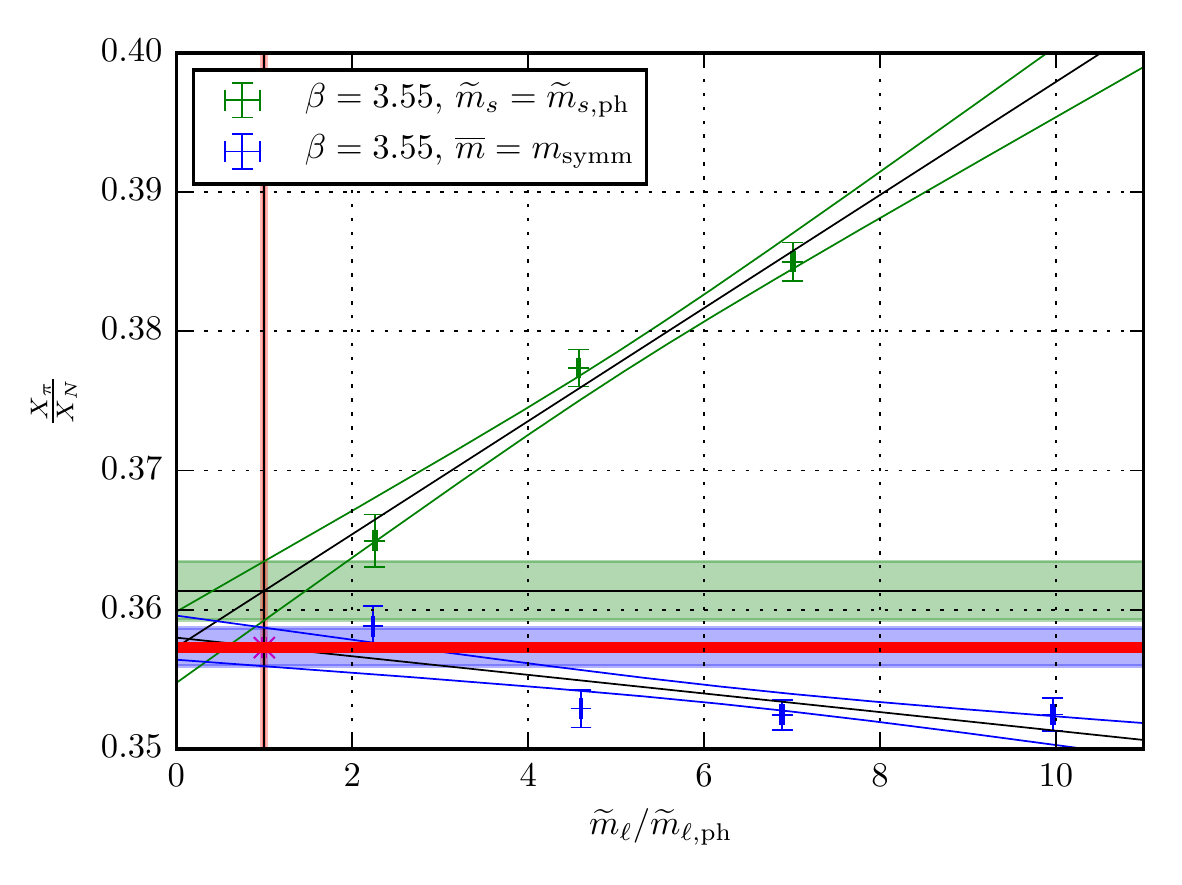}
     }
\caption{\label{fig:X}Ratios of average hadron masses for $\beta=3.4$ (left) and $\beta=3.55$ (right) along the two chiral trajectories
  $\overline m = m_\mathrm{symm}$ (blue) and $\widetilde{m}_s=\widetilde{m}_{s,\mathrm{ph}}$ (green).
  The red bands correspond to the physical values. The horizontal bands
to the values extrapolated to the physical point.}
\end{figure}

We now proceed to baryon masses.
We define the average octet and decuplet baryon masses
$X_N$ and $X_{\Delta}$ as
$X_N \equiv \left( m_N + m_\Sigma + m_\Xi  \right)/3$ and $X_{\Delta}=(2M_{\Delta}+M_{\Omega})/3$, respectively~\cite{Bietenholz:2011qq}. These, to leading order
in SU(3) $\chi$PT are constant along
the $\overline{m}=m_{\rm symm}$ trajectory while individual baryon masses
are linear functions of the quark masses. Since along our
$\overline{m}=m_{\rm symm}$ trajectory the strange quark mass linearly
depends on the light quark mass (up to lattice artefacts),
as a first approximation one can attempt linear fits.

In Fig.~\ref{fig:oct} we plot the octet and decuplet baryon masses
in a similar way as the pseudoscalar mesons discussed above.
Again, the independent linear fits describe all data reasonably well.
In particular, we find agreement of both chiral trajectories at
the physical point. For comparison we also show the experimental
values (red lines).
Finally, in Fig.~\ref{fig:X} we extrapolate ratios
of average masses. Also in this case at the physical point
we find agreement on the
one- to two-sigma level between the two mass plane trajectories and
experiment.

For all these quantities we are currently investigating
lattice spacing effects and effects from higher orders
in chiral perturbation theory. 

\section{Scale setting}
\label{sec:scale}

\begin{figure}[t]
\centerline{
    \includegraphics[width = .5\textwidth]{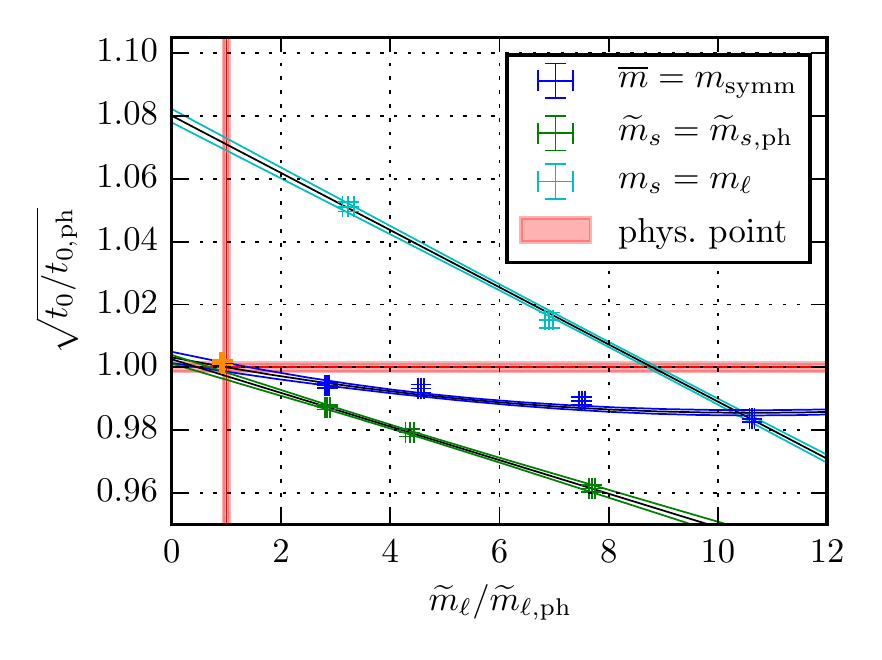}
    \includegraphics[width = .5\textwidth]{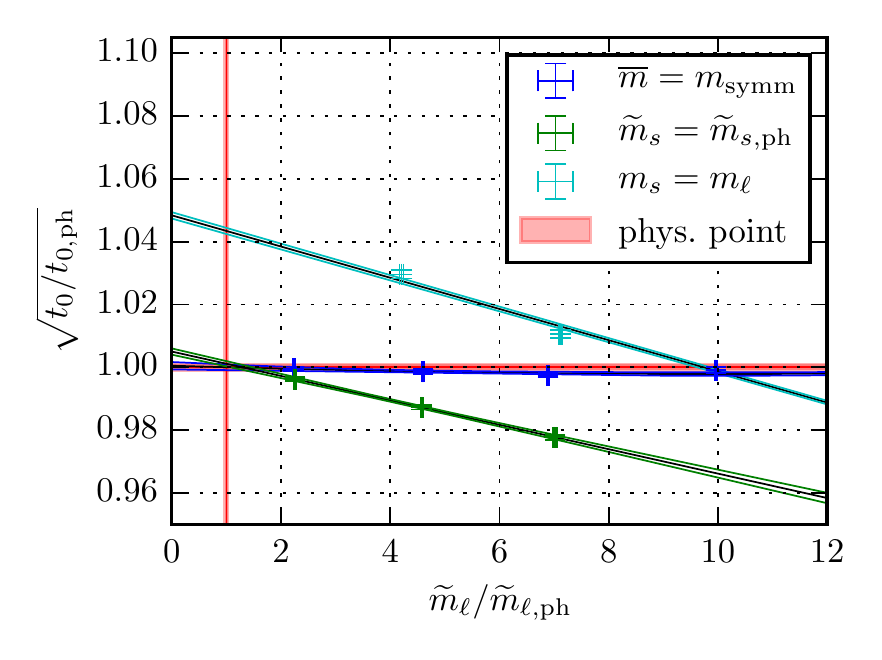}}
    \caption{\label{fig:t0}Chiral extrapolations of $t_0$ for $\beta=3.4$ (left) and $\beta=3.55$ (right) along the two chiral trajectories
  $\overline m = m_\mathrm{symm}$ (blue) and $\widetilde{m}_s=\widetilde{m}_{s,\mathrm{ph}}$ (green).
  The red bands correspond to the physical values.  Plots taken from Ref.~\cite{Bali:2016umi}.}
\end{figure}

To assign a scale to the lattice spacing at a given value
of the inverse coupling $\beta$ we need to equate a dimensionful observable to
its experimental value. The latter is only available at the physical
point, thereby necessitating a chiral extrapolation. Purely gluonic observables
have a reduced quark mass dependence and therefore a milder chiral
extrapolation. Some of these can also be determined quite accurately with a small
computational effort and therefore, while not measurable in experiment,
are a prime choice for an intermediate scale to translate between
different lattice spacings. One such example is $t_0$. As can be seen in
Fig.~\ref{fig:t0} its mass dependence is very mild along
$\overline m = m_\mathrm{symm}$, nevertheless there
is some curvature visible in the data, especially at $\beta=3.4$.
However, in the end one has to assign a continuum, physical point value
to $t_0$. Here we have used the value obtained by BMW-c~\cite{Borsanyi:2012zs}
using the mass of the $\Omega$ baryon. Their error on $t_0$ translates
into a 1.7\% error on the lattice spacing.
Another (compatible) result was obtained recently on CLS ensembles from
pseudoscalar decay constants~\cite{Bruno:2016plf}.
\begin{figure}[b]
    \centerline{
    \includegraphics[width = .33\textwidth]{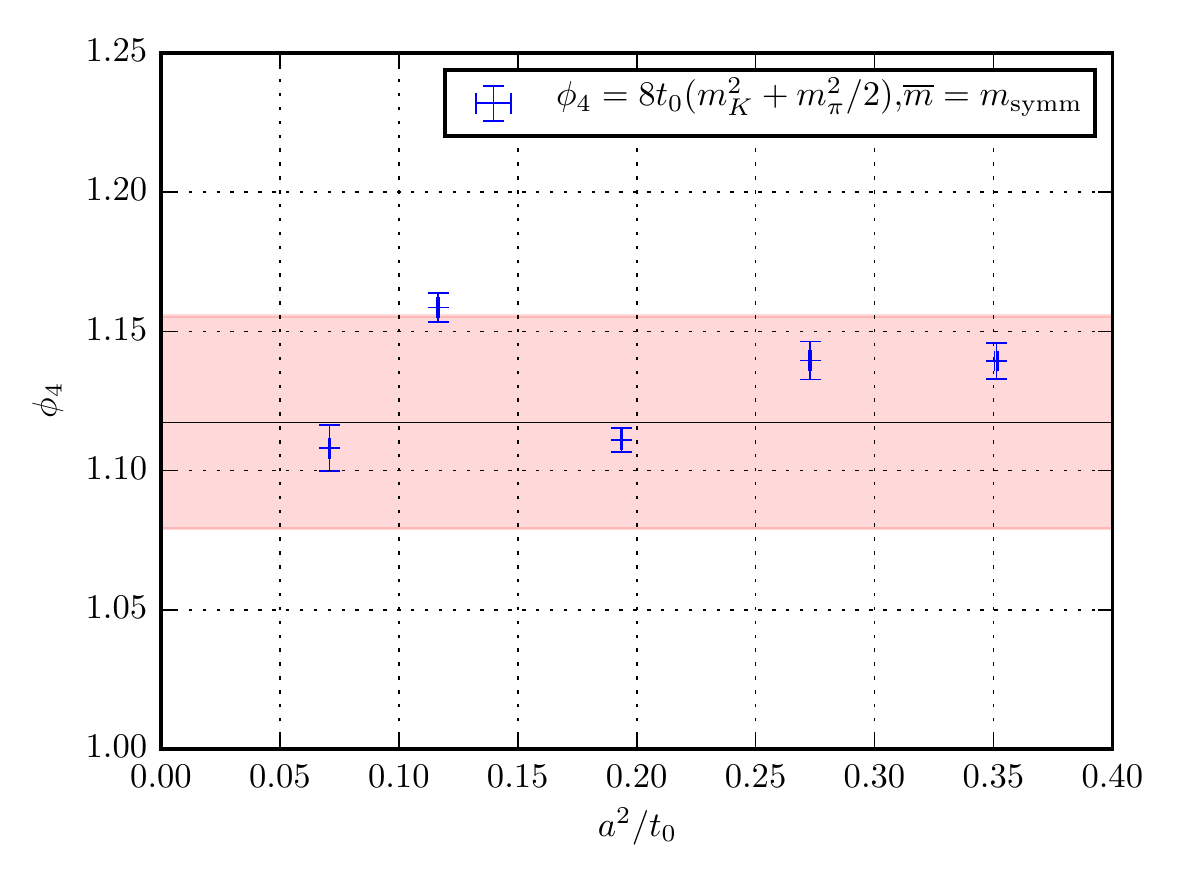}
    \includegraphics[width = .33\textwidth]{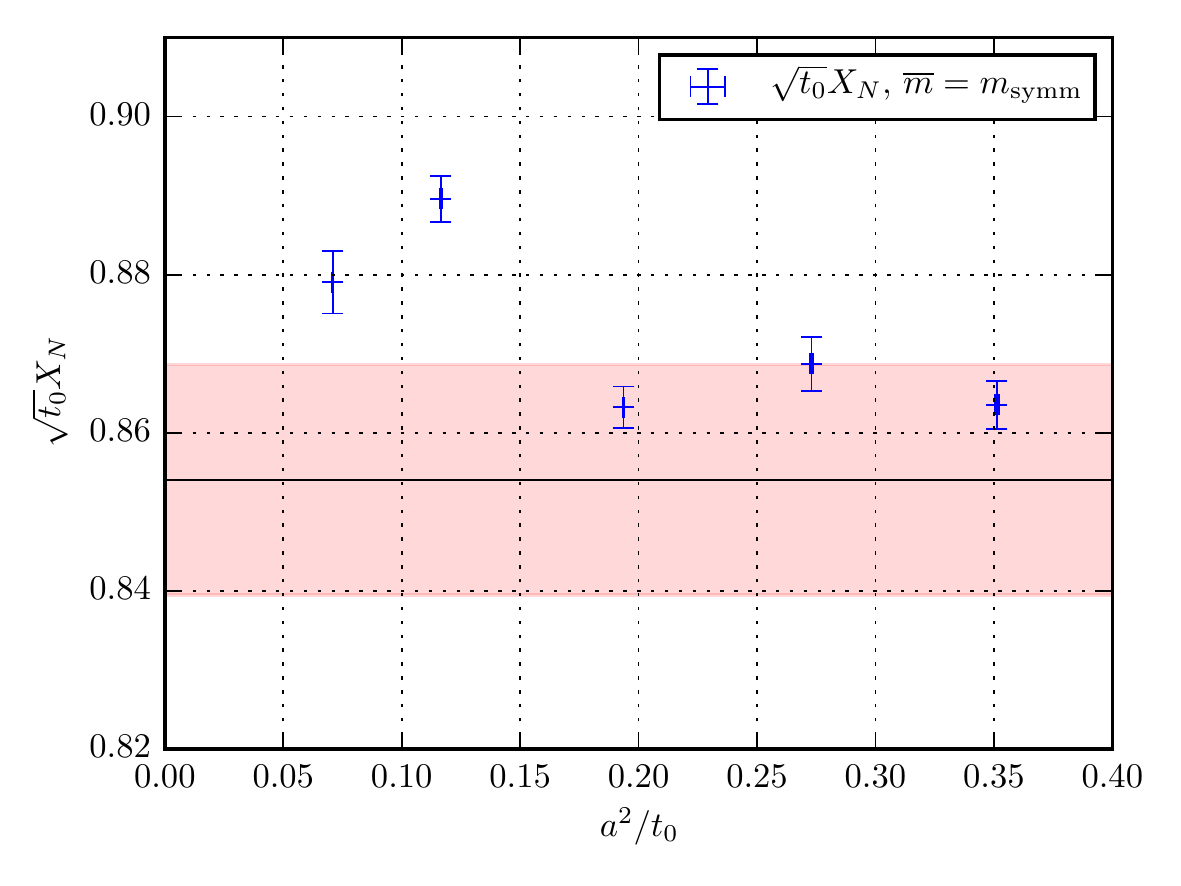}
    \includegraphics[width = .33\textwidth]{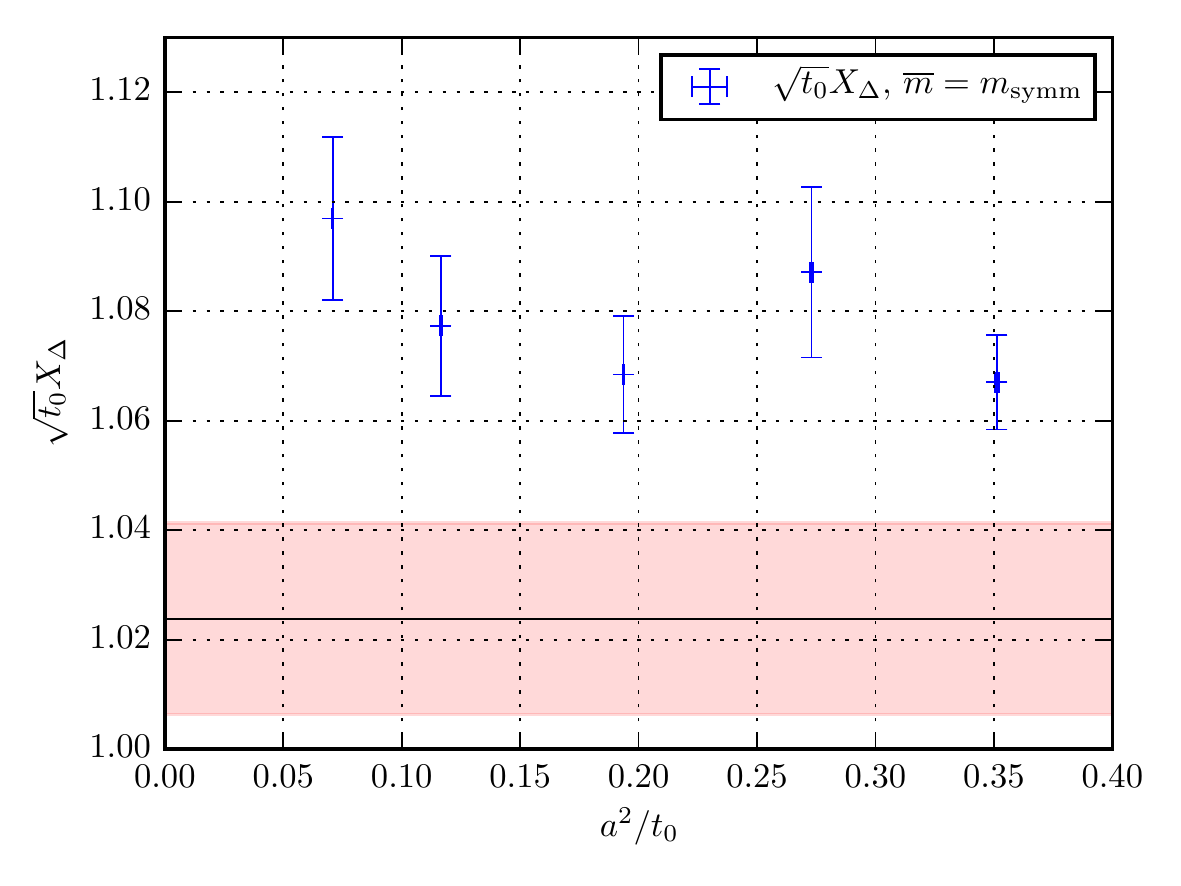}
    }
    \caption{\label{fig:contsym}$\phi_4$, $\sqrt{t_0} X_N$, and$\sqrt{t_0} X_\Delta$ at the symmetric point versus $a^2/t_0$. 
    The red bands correspond to the physical values.}
\end{figure}

One may also use continuum limit extrapolated baryon masses to
set the scale.
Below we present a comparison of relative errors
that we get when determining the scale
using different baryons:
\begin{equation}
m_\Xi\,:\, 0.4\%\,,\,\, m_{\Xi^\star} \,:\, 0.9\%\,\,\,
m_\Omega\,:\, 0.7\%\,,\,\, X_N\,:\, 0.6\%.
\end{equation}
These errors are all smaller than 1.7\%, however, we have not
yet carried out the continuum limit extrapolation. Also these values
must be considered preliminary until we scrutinize the chiral extrapolations.
To illustrate the effect on the lattice spacings of using different
scale setting methods we quote (preliminary) numbers for $a$ at
$\beta=3.4: a_{X_N} \approx 0.0833(4)\text{fm}$, compared to $ a_{t_0} = 0.0854(15)\text{fm}$, and at
$\beta=3.55: a_{X_N} \approx 0.0632(5)\text{fm}$, compared to $ a_{t_0} = 0.0644(11)\text{fm}.$
      
\section{Continuum extrapolation along the symmetric point}
\label{sec:cont}
In the near future we will perform a combined extrapolation of the
hadron spectrum
to the continuum limit and the physical point. It is particularly
subtle to disentangle $\mathcal O(a^2)$ effects from quark mass
effects. Therefore, as a first step it is informative to investigate the
continuum limit of various quantities at the symmetric point, which
by definition is at a fixed renormalized quark mass since the dimensionless
combination $\phi_4$ is matched across the different lattice spacings.
However, there is some deviation from the initial
target value $\phi_4=1.15$ between different values of $\beta$, which is
illustrated in the left plot of Fig.~\ref{fig:contsym} and is also
visible in Fig.~\ref{fig:simdetail}. 
From Fig.~\ref{fig:contsym} it is clear that this also affects other
quantities.

\begin{figure}[t]
\centerline{
    \includegraphics[width = .5\textwidth]{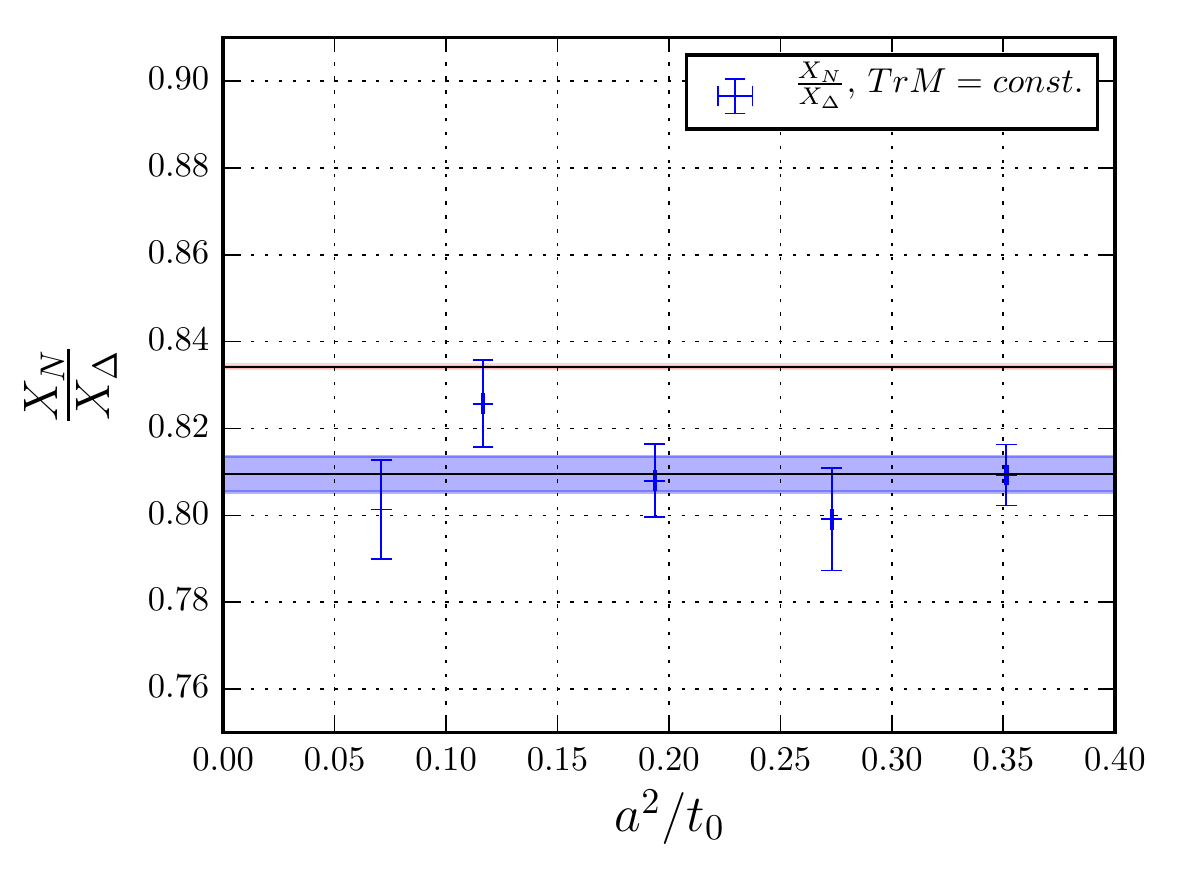}
    \includegraphics[width = .5\textwidth]{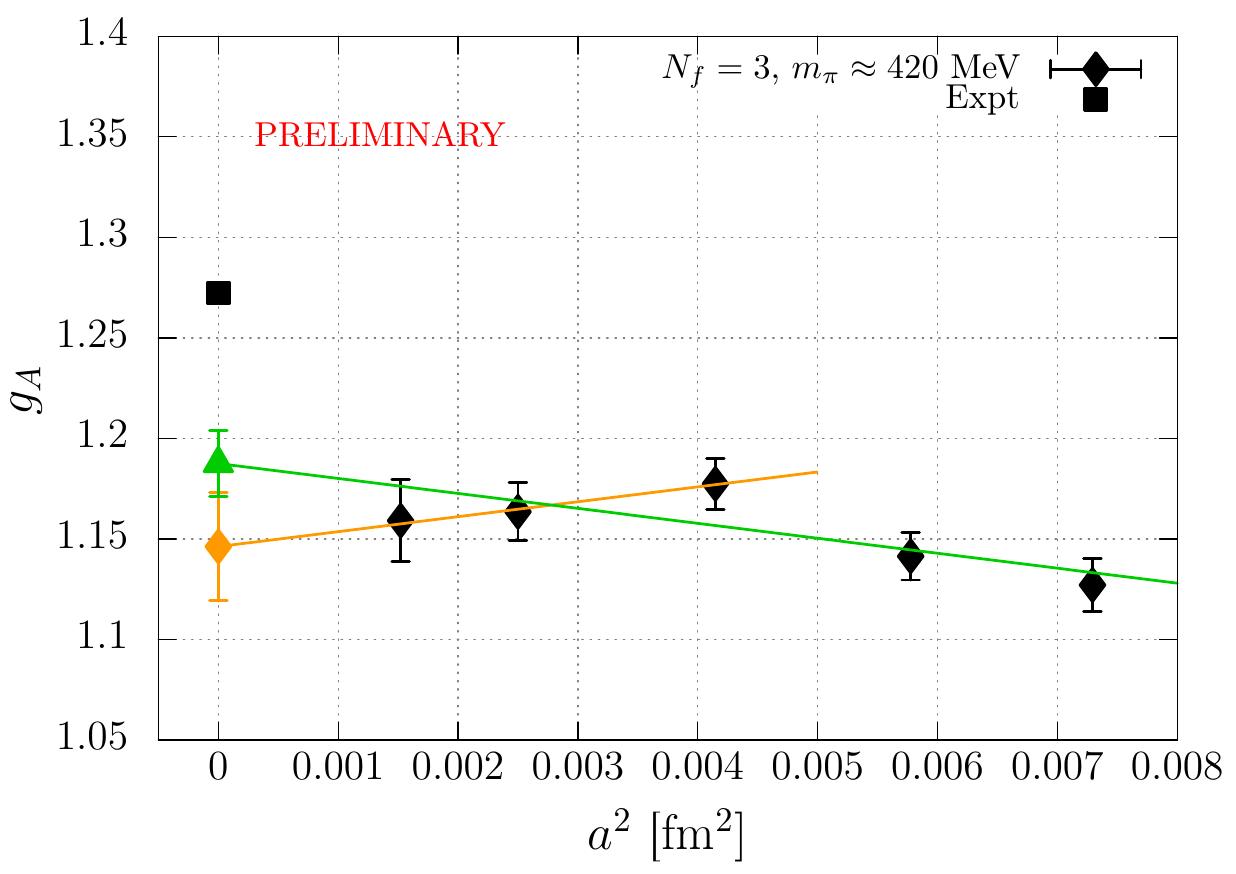}}
\caption{\label{fig:Xcont} $X_N/X_\Delta$ (left plot) at the symmetric point versus $a^2/t_0$. The red band corresponds to the physical value, the blue band indicates the error from a fit to a constant.
Right: $g_A$ at the symmetric point versus $a^2$. The lines correspond to linear fits in $a^2$. The left-most four data points are also compatible
with a constant.}
\end{figure}

In order to take the continuum limit, usually we will have to correct for
this mismatch. Here we focus on two quantities which
we expect to exhibit only a very small quark mass dependence
along the symmetric line. In Fig.~\ref{fig:Xcont} we show the ratio of
the average octet over decuplet mass as a function of $a^2$ in units of $t_0$.
We observe a very flat behaviour suggesting a rather mild cut-off
dependence for this observable.

Another quantity which should not significantly change, varying the
pion mass by a few per cent around 420~MeV, is the axial isovector charge
of the nucleon $g_A$. This was determined following the
$N_f=2$ computation described in Ref.~\cite{Bali:2014nma}, using the renormalization
constant $Z_A$ of Ref.~\cite{Bulava:2016ktf} and the improvement
parameter $b_A$ of Ref.~\cite{Korcyl:2016ugy}.
In the right panel of Fig.~\ref{fig:Xcont} we plot $g_A$, again at the
symmetric point, together with two fits linear in $a^2$. Clearly, we would have
overestimated the correct continuum limit value had we only fitted the
coarsest three data points (fit not shown). This demonstrates that
a broad window of lattice spacings is compulsory. Note that we
significantly underestimate the experimental physical point value,
as we should at such a large quark mass.
A more detailed analysis is ongoing.

\section{Summary and Outlook}
We have presented results on AWI quark masses, pseudoscalar meson and baryon
spectra as well as on $g_A$
from lattice simulations on $N_f=2+1$ ensembles generated
within CLS. The use of open boundaries avoids topological freezing
as $a\to 0$ and will allow us to take a controlled continuum limit.
This was demonstrated for two examples, namely the ratio
of the nucleon over the $\Delta$ mass and the axial
coupling of the nucleon $g_A$, albeit at a large pion mass value.
Our results cover pion masses
$m_{\rm PS} \approx 200$--$700 \MeV$ at lattice spacings
ranging from $a\approx 0.085$ down to $a\approx 0.04 \fm$.
Using LO $\chi$PT, we have extrapolated the masses
to the physical point along two trajectories
($\overline m = m_\mathrm{symm}$ and 
$\widetilde{m}_s=\widetilde{m}_{s,\mathrm{ph}}$).
Combined fits using Gell-Mann--Okubo type expansions as
well as $\SU(3)$ $\chi$PT
along all three trajectories (including $m_s = m_{\ell}$)
are currently in progress.
This will allow us to extract $\SU(3)$ low energy constants (LECs),
while the $\widetilde{m}_s=\widetilde{m}_{s,\mathrm{ph}}$ trajectory
yields additional information on the $\SU(2)$ LECs.
    
The spectrum calculations constitute a preparatory step for an independent
determination of the lattice scale and of light quark masses.
A more detailed analysis of the nucleon structure and other additional
observables covering a large range of lattice spacings
will follow in the near future.

\bibliography{lat16v2}{}
\bibliographystyle{JHEP_mod}

\end{document}